\documentclass[10pt,twoside,a4paper,twocolumn]{article}
\pdfoutput=1
\usepackage{amsmath,amsfonts,amssymb}
\usepackage{graphicx}
\usepackage{authblk}
\usepackage{subcaption}
\usepackage{hyperref}
\usepackage{cite}

\begin{document}

\title{\vspace{-4.5cm}Spin coherent quantum transport of electrons between defects in diamond}
\author[a]{L. M. Oberg}
\author[a]{E. Huang}
\author[a]{P. M. Reddy}
\author[b]{A. Alkauskas}
\author[c]{A. D. Greentree}
\author[d]{J. H. Cole}
\author[a]{N. B. Manson}
\author[e]{C. A. Meriles}
\author[a]{M. W. Doherty}
\affil[a]{Laser Physics Center, Research School of Physics and Engineering, Australian National University, Australian Capital Territory 2601, Australia}
\affil[b]{Center for Physical Sciences and Technology, Vilnius LT-10257, Lithuania}
\affil[c]{ARC Center of Excellence for Nanoscale BioPhotonics, School of Science, RMIT University, Melbourne, VIC, 3001, Australia}
\affil[d]{Chemical and Quantum Physics, School of Science, RMIT University, Melbourne, Victoria 3001, Australia}
\affil[e]{Department of Physics, CUNY-City College of New York, New York, New York 10031, USA}

\maketitle
\begin{abstract}
The nitrogen-vacancy color center in diamond has rapidly emerged as an important solid-state system for quantum information processing. While individual spin registers have been used to implement small-scale diamond quantum computing, the realization of a large-scale device requires development of an on-chip quantum bus for transporting information between distant qubits. Here we propose a method for coherent quantum transport of an electron and its spin state between distant NV centers. Transport is achieved by the implementation of spatial stimulated adiabatic Raman passage through the optical control of the NV center charge states and the confined conduction states of a diamond nanostructure. Our models show that for two NV centers in a diamond nanowire, high fidelity transport can be achieved over distances of order hundreds of nanometres in timescales of order hundreds of nanoseconds. Spatial adiabatic passage is therefore a promising option for realizing an on-chip spin quantum bus.
\end{abstract}

\section*{Introduction}
Diamond-based quantum information processing (QIP) is possible due to the remarkable properties of the nitrogen-vacancy (NV) defect center. The spin-states of individual NV centers can be optically addressed for initialization and read-out\cite{Doherty2013} and possess the longest room-temperature spin-coherence time of any solid-state defect\cite{Balasubramanian2009}. Cryogenic temperatures further enhance these spin properties by up to four orders of magnitude\cite{Jarmola2012} and enable spin-photon entanglement through coherent optical transitions\cite{Togan2010}. This includes adiabatic control of the NV spin using optical pulses\cite{Yale2013}. These properties have been used to operate individual spin registers for small-scale quantum computing including demonstrations of error correction\cite{Taminiau2014}, simple algorithms\cite{Shi2010} and simulations\cite{Wang2015}. Spin registers have been realized through clusters of NV centers as well as single NV centers coordinated with clusters of paramagnetic defects such as ${}^{13}$C atoms and substitutional nitrogen ($\text{N}_\text{S}$) centers\cite{Gaebel2006,Hanson2006,Neumann2008,Neumann2010,Dolde2013,Dolde2014,Waldherr2014a}.

Individual spin registers are ultimately limited by the number paramagnetic defects addressable by a single NV center (5-10 strongly coupled qubits\cite{Doherty2017}). Development of large-scale QIP therefore requires the use of quantum buses to entangle a network of interconnected clusters. At cryogenic temperatures optical quantum buses have been used to entangle spin registers separated by macroscopic distances\cite{Togan2010,Bernien2013,Wehner2015}. Unfortunately this option is not viable for scalable QIP as optical buses cannot be compactly incorporated on chip. One scalable suggestion is intra-cluster spin transport along a chain of paramagnetic defects\cite{Yao2012a,Nikolopoulos2014}. Despite the promise of room-temperature operation, engineering the spin chain is difficult with existing fabrication techniques. Recently, Doherty \textit{et al}\cite{Doherty2016} have proposed entanglement using semi-classical electron transport between two NV-${}^{14}\text{N}_\text{S}$ pairs embedded in a diamond nanowire. This scheme is severely limited as semi-classical transport is prone to erroneous capture from $\text{N}_\text{S}^+$ defects, surface traps\cite{Stacey2019} or surface emission\cite{Kaviani2014a}.

Consequently, here we propose an on-chip spin quantum bus for diamond QIP using spatial stimulated adiabatic Raman passage (STIRAP). Spatial STIRAP involves the transport of massive particles between spatially distinct locations using conventional STIRAP type coupling. It was originally explored in the context of electronic transfer in double quantum dots\cite{Brandes2002,Vitanov2017} and for general methods of spatial adiabatic passage which promise robust quantum information transport\cite{Greentree2004,Menchon-Enrich2016}. The conventional STIRAP technique is a method of population transfer between the two lowest levels of a $\Lambda$ type atomic system\cite{Bergmann2002}. By applying optically coupling fields in the so-called counter intuitive direction, where the unpopulated transition is coupled before the populated transition, the system can be maintained in the optical dark state, which suppresses population in the central (excited) state. This suppression of population in the central state, which will ideally have zero population in the adiabatic limit, leads to the suppression of the effects of spontaneous emission. In the context of NV centers, STIRAP has been proposed\cite{Coto2017a} and demonstrated\cite{Zhou2017} for single qubit control, however we are not aware of any proposals for the spatial transport of electrons between NV centers of the form considered here.

\begin{figure}[]
    \centering
    \includegraphics[width=0.35\textwidth]{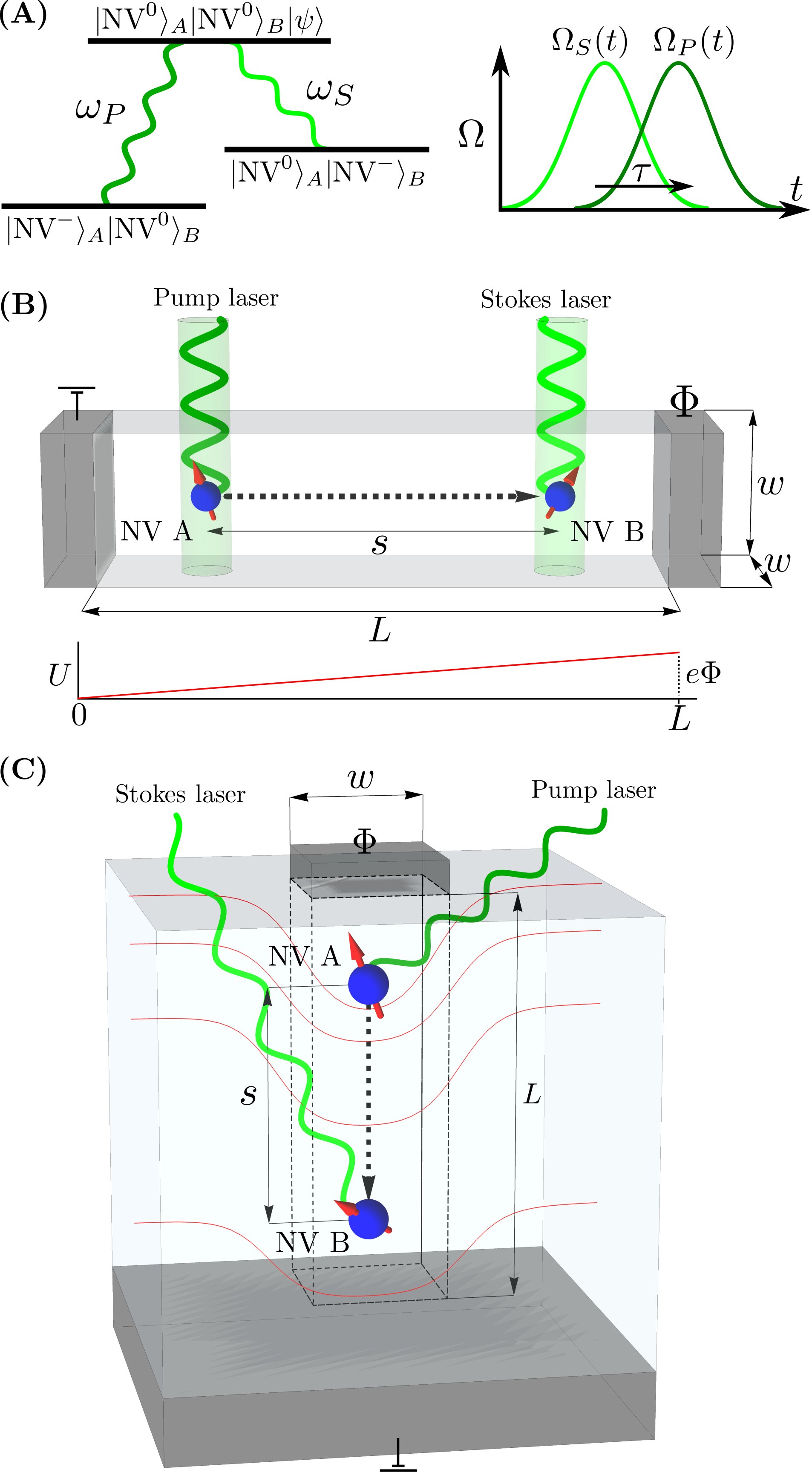}
    \caption{\textbf{(A)} $\Lambda$-level scheme for two NV centers in a diamond nanowire. Adiabatic passage is mediated by mutual coupling to the conduction band minimum state, $|\psi\rangle$. Applying a Stokes laser pulse before a pump laser pulse with some overlapping time $\tau$ results in adiabatic quantum transport of the electron and its spin-state from NV $A$ to NV $B$. \textbf{(B)} Two NV centers labeled $A$ and $B$ are separated by a distance $s$ in a surface confined nanowire of length $L$ and width $w$. A potential difference of $\Phi$ is applied to electrode plates affixed to either end of the wire. This generates a gradient electric potential (graphed as $U$) which lifts the degeneracy of states $|1\rangle$ and $|3\rangle$ in the STIRAP level scheme. To avoid erroneous excitations to higher conduction band states, the pump and Stokes lasers must individually address each NV center and therefore be spatially separated. \textbf{(C)} Schematic of an electrostatically confined nanowire. The two NV centers are aligned co-axial and perpendicular to the surface of a diamond substrate. A square nano-electrode of width $w$ is affixed to the surface above the NV centers while a grounded electrode plate is affixed to the opposite side of the substrate. Applying a small positive voltage $\Phi$ to the nano-electrode generates a nanowire confining potential (red equi-potentials) which extends an effective depth $L$ into the substrate.}
    \label{fig:wires}
\end{figure}

In our proposal for spatial STIRAP the two disconnected states of the $\Lambda$ scheme correspond to an electron occupying one of two distant NV centers embedded in a diamond nanostructure. As depicted in Figure~\ref{fig:wires}~\textbf{(A)}, we envision mediating adiabatic passage through states of the conduction band minimum, which are discretized by the confining potential of the nanostructure. While all nanostructures lead to discretization, we will consider nanowires as an archetypal structure. We have identified two different designs for realizing a nanowire confining potential, via the potential of a diamond surface or an electrostatic potential applied by electrodes in a bulk structure. Schematics for the transport scheme in these designs are presented in Figure~\ref{fig:wires}. While STIRAP is performed similarly in both cases, the susceptibility to different decoherence mechanisms varies. We also predict that quantum transport will maintain coherence of the electron's spin. Spin transport is known theoretically to be excellent in diamond due to an indirect band gap, low spin-orbit coupling, low electron-phonon (e-p) scattering and low background impurity spins\cite{Restrepo2012}. Here, under STIRAP we explicitly exploit the low background impurities and low spin-orbit coupling of the conduction band to achieve spin coherent transport.

This proposal for quantum transport is fundamentally different from existing schemes for spatial STIRAP. As coupling is mediated optically rather than through manipulation of tunneling amplitudes\cite{Menchon-Enrich2016}, the spatially separated states are truly disconnected and there is no occupation of the intermediary space during transport. In this sense, it is colloquially akin to teleportation of a massive particle. Furthermore, to our knowledge this is the first spatial STIRAP proposal which includes coherent spin transport.

In Section~\ref{sec:transport} we present a proposal for implementing STIRAP in a diamond nanowire including a spin initialization and measurement protocol. These spin protocols play the essential role of identifying the transported electrons (thereby distinguishing them from background sources) and will be the means for which spin entanglement is mediated. We identify a feasibility condition for transport, $\Delta E_c/\hbar \gg \Omega \gg \Gamma$, where $\Delta E_c$ defines the energy gap between the first and second conduction states, $\Omega$ is the Rabi frequency of the optical transition between defect and conduction states, and $\Gamma$ is the optical decoherence rate. As a conservative estimate, these constraints can be considered satisfied when $\Delta E_c/\hbar \gtrsim 10\Omega \gtrsim 10\Gamma$. In Section~\ref{sec:confinement} we calculate $\Delta E_c$ by modeling the effect of the nanowire confining potential on the diamond conduction band. Effective mass theory is used to derive a level scheme and determine energies of discretized conduction states. In section~\ref{sec:optical} we present the first \textit{ab-initio} calculations for the photoionization mechanism of the NV center. This is used to evaluate values of $\Omega$ that can be practically achieved. We then identify optical decoherence mechanisms that contribute to $\Gamma$. Finally, in Section~\ref{sec:optimization} we evaluate $\Omega$, $\Delta E_c$ and $\Gamma$ at cryogenic temperatures for both of the nanowire designs introduced previously. The dimensions of the wires are optimized to maximize the Rabi frequency relative to decoherence and to satisfy the STIRAP feasibility constraints.

\section{Quantum transport in diamond nanowires}
\label{sec:transport}

STIRAP is a method for evolving a three-state system from an initial state $|1\rangle$ to a final state $|3\rangle$\cite{Bergmann2002,Ivanov2004}. This is achieved through coupling to a mutual intermediary state, $|2\rangle$, even though the direct transition $|1\rangle\rightarrow|3\rangle$ may be forbidden. To do so, a pump laser pulse with Rabi frequency $\Omega_P(t)$ is tuned to the $|1\rangle\rightarrow|2\rangle$ transition while a Stokes laser pulse with Rabi frequency $\Omega_S(t)$ is tuned to the $|2\rangle\rightarrow|3\rangle$ transition. Applying the Stokes laser pulse before the pump laser pulse with some overlapping time $\tau$ results in an adiabatic evolution of the state from $|1\rangle\rightarrow|3\rangle$ without any population of $|2\rangle$. The requirement for adiabaticity is given by $\Omega\tau \gg 1$, where $\Omega^2=\Omega_P^2 + \Omega_S^2$ is the effective Rabi frequency.

\begin{figure*}[]
	\centering
	\includegraphics[width=0.65\textwidth]{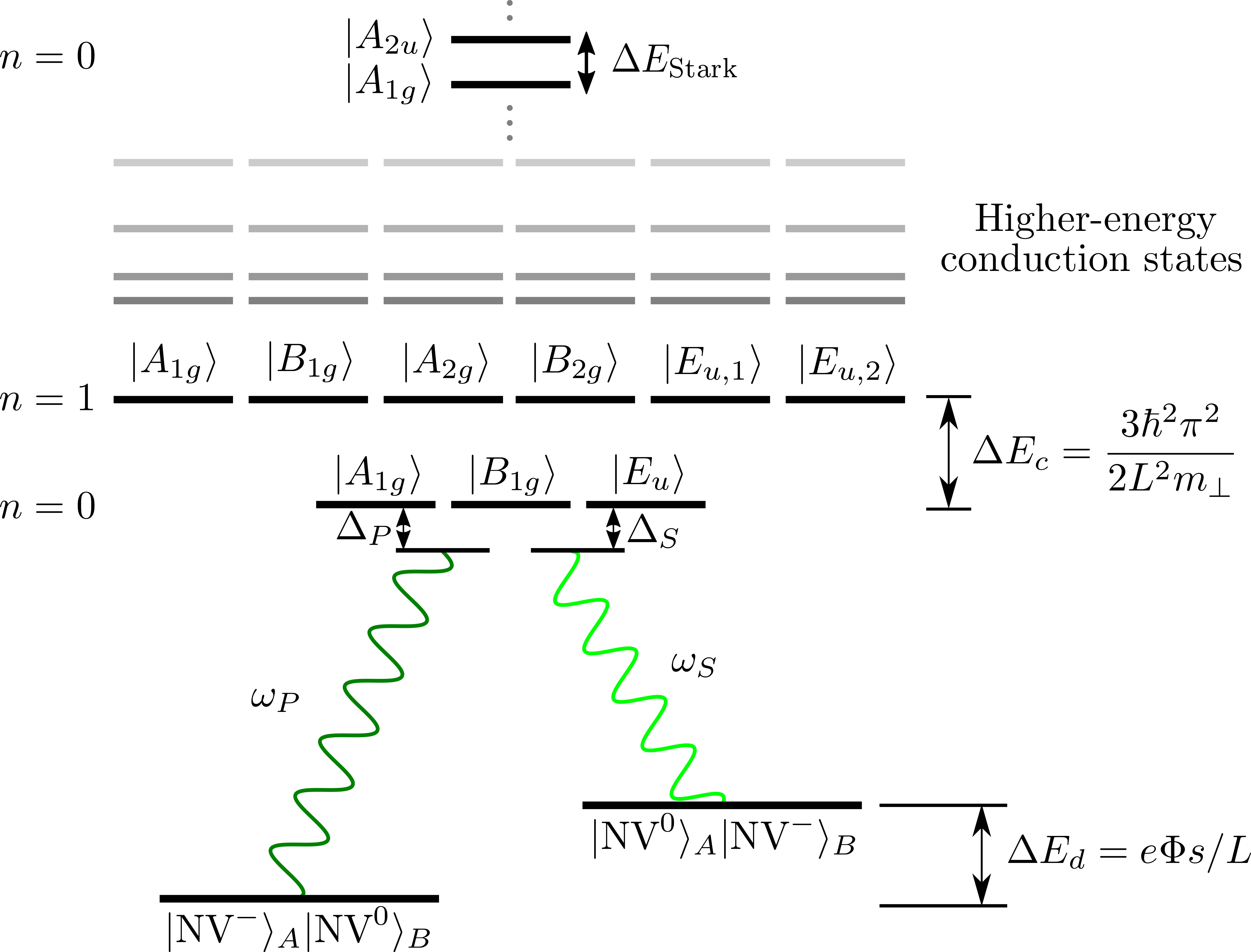}
	\caption{Nanowire energy level diagram for the STIRAP transport protocol. The electron is transported from the initial defect state $|\text{NV}^-\rangle_A|\text{NV}^0\rangle_B$ to the target defect state $|\text{NV}^0\rangle_A|\text{NV}^-\rangle_B$ using laser pulses with Rabi frequencies $\Omega_P$ and $\Omega_S$ to mediate adiabatic passage with the four-fold degenerate conduction band minima states. Implicit in the labeling of conduction band states is the combined electronic state of the defects, $|\text{NV}^0\rangle_A|\text{NV}^0\rangle_B$. Not to scale.}
	\label{fig:level_scheme}
\end{figure*}

We propose using STIRAP for quantum transport of an electron between two NV centers. As presented in Figure~\ref{fig:wires}, the NVs are embedded within the nanowire confining potential and labeled as $A$ and $B$. This allows us to define a three-level system given by
\begin{align*}
|1\rangle &= |\text{NV}^-\rangle_A|\text{NV}^0\rangle_B \\
|2\rangle &= |\text{NV}^0\rangle_A|\text{NV}^0\rangle_B|\psi\rangle \\
|3\rangle &= |\text{NV}^0\rangle_A|\text{NV}^-\rangle_B,
\end{align*}
where $|\text{NV}^-\rangle$ and $|\text{NV}^0\rangle$ correspond to the ${}^3\text{A}_2$ and ${}^2\text{E}$ electronic ground states of the negative and neutral charge states of the NV centers and $|\psi\rangle$ is a nanowire conduction band minimum state. A schematic of the STIRAP level scheme is presented in Figure~\ref{fig:level_scheme}. Adiabatic passage between states $|1\rangle$ and $|3\rangle$ therefore constitutes electron transport from NV $A$ to NV $B$ with no occupation of the intermediary space. The degeneracy of states $|1\rangle$ and $|3\rangle$ may be lifted by applying a potential difference $\Phi$ between each end of the length $L$ wire. For a defect separation distance $s$, this induces an energy splitting of $\Delta E_d=e\Phi s/L$.

Let us now consider transportation of the electron spin during STIRAP. This requires a protocol for spin initialization of both $|\text{NV}^-\rangle_A$ and $|\text{NV}^0\rangle_B$ which we now present in turn. Firstly, suppose we wish to transport the spin-state $|\uparrow\rangle$. We restrict our attention to photoionization by adsorption of a single photon from an optical pulse and do not consider the two-photon ionization process of the $\text{NV}^-$\cite{Aslam2013}. As the ionizing pulse cannot differentiate between the two electrons occupying the $e$-orbitals of $|\text{NV}^-\rangle_A$, both must be initialized into $|\uparrow\rangle$. This can be achieved through optical polarization of $|\text{NV}^-\rangle_A$ into the $m_s = 0$ triplet state followed by application of a microwave control pulse. The resulting spin-state can be expressed in the two $e$-orbital basis as $|+1\rangle=|\uparrow\rangle|\uparrow\rangle$. Hence, through this protocol we can guarantee ionization of a spin-up electron during STIRAP.

Now consider initialization of the $|\text{NV}^0\rangle_B$ spin-state. Post STIRAP, electron transportation can only be validated if read-out of the $|\text{NV}^-\rangle_B$ spin-state agrees with the spin initialized at $|\text{NV}^-\rangle_A$. Otherwise, it is impossible to discern if charging of $|\text{NV}^0\rangle_B$ into $|\text{NV}^-\rangle_B$ was instead due to erroneous electron capture. It is therefore essential for read-out that $|\text{NV}^0\rangle_B$ be initialized into either the $|\uparrow\rangle$ or $|\downarrow\rangle$ spin eigenstates. To see how this could be achieved, consider that the $|\text{NV}^-\rangle_B$ spin-state after transport (assuming $|\text{NV}^0\rangle_B$ was initialized as $|\uparrow\rangle$) is given by $|\uparrow\rangle|\uparrow\rangle=|+1\rangle$. Similarly, we obtain the $m_s=0$ triplet-state when $|\text{NV}^0\rangle_B$ is initialized as $|\downarrow\rangle$. Transport can then be confirmed by performing read-out of the spin-state at NV $B$.

The spin initialization of NV $B$ can be achieved as follows. Consider NV $B$ initially prepared into the negative charge state with spin projection $m_s = +1$. Application of a $>2.6$~eV photoionizing pulse will then eject an electron with $|\uparrow\rangle$ spin, leaving NV $B$ in the neutral charge state with $|\uparrow\rangle$ spin. Equivalently, the $|\downarrow\rangle$ spin-state can be prepared for NV $B$ prepared in the $m_s = -1$ state. While not proven, it is assumed that photoionization is a spin-conserving process. Even if this is not the case, one way to ameliorate the risk of non-conserving spin transitions is to polarize the N nuclear spin when NV $B$ is in the negative charge state. After the ionizing pulse, this nuclear spin polarisation can be swapped back onto the electron spin of the neutral charge state via microwave control. Such a scheme is now feasible given recent progress has identified the $\text{NV}^0$ spin Hamiltonian\cite{Barson2019}. 

Transportation also requires that $\text{NV}^0$ spin-relaxation is long with respect to the STIRAP operation time. As of yet the $\text{NV}^0$ ground state has not been directly observed through electron paramagnetic resonance. Whilst this would presumably imply a high rate of spin relaxation\cite{Felton2008}, new work by Barson \textit{et al} indicates that this is likely not the case. Rather, the apparent absence of EPR is due to strain broadening in ensembles\cite{Barson2019}. Furthermore, due to the electronic similarities between the $\text{NV}^0$ and $\text{Si-V}^-$ center\cite{Jahnke2015}, e-p scattering is expected to cause spin dephasing but not relaxation of $\text{NV}^0$. Consequently, at cryogenic temperatures $\text{NV}$ $B$ should maintain its spin projection throughout transport.

During transport, spin decoherence mechanisms can be neglected. This remains true as long as the spin-quantization axis of the conduction states match those of the NV center and there is no rapid spin dephasing of these conduction states. This is investigated in supplementary section~B, where we present symmetry arguments within the closure approximation. We prove that if the NV is co-axial to the wire then the spin-quantization axis is conserved throughout STIRAP. Conveniently, NV centers can be preferentially aligned with a (111) wire axis during fabrication with chemical vapor deposition\cite{Michl2014}. For other nanowire directions, the spin-quantization axis of the conduction state can be aligned with the NV center via the application of a magnetic field. The magnitude of this field must be sufficiently large relative to the component of the spin-orbit interaction orthogonal to (111).

There are two further requirements for successful implementation of STIRAP. Firstly, coherent transport demands there be no off-resonant coupling to states other than the conduction band minimum. Hence, the energy splitting between the first and second conduction states $\Delta E_c$ must be significantly greater than the Rabi frequency, $\Delta E_c/\hbar\gg\Omega$. Secondly, the optical decoherence rate $\Gamma$ sets a bound on $\Omega$ since it induces an unknown phase on the electron wave function as the transition energy fluctuates. Provided the Rabi frequency is large ($\Omega\gg\Gamma$), this dephasing will be sufficiently small to achieve high fidelity transport. The following two sections address each of these feasibility requirements in turn. We first investigate the effects of wire confinement on the conduction band before evaluating the Rabi frequency and identifying sources of optical decoherence.

\section{Diamond nanowire conduction band}
\label{sec:confinement}

The confining potential of the nanowire leads to discretization of the conduction band suitable for STIRAP. These confinement effects are well described through effective mass theory (EMT)\cite{Stoneham1976}, as detailed in supplementary section~A. As shown in Figure~\ref{fig:wires}, the nanowire was modeled as a square prism of mono-crystalline diamond with dimensions $w\times w\times L$ ($w<L$). The two NV centers are separated by a distance $s$ and located on the longitudinal. Consider that bulk diamond possesses six equivalent conduction band minima, located at the $k$-points $\vec{K}_i$ between the $\Gamma$ and $X$ points in each orthogonal axis of the fcc Brillouin zone. The dispersion relationship at each $\vec{K}_i$ is modeled using diamond's anisotropic mass tensor and is referred to as a valley\cite{Madelung2004a}.

In the effective mass framework, the $n^\text{th}$ conduction band state for the $i^\text{th}$ valley of the nanowire is given by
\begin{equation}\label{wav}
\psi_{n,i}(\vec{r})=F_{n,i}(\vec{r})e^{i\vec{K}_i\cdot \vec{r}}u_i(\vec{r}),
\end{equation}
where $F_{n,i}$ is an envelope function determined by the confining potential, and $u_i$ is the bulk-diamond Bloch function. For simplicity we approximate the confining potential as an infinite square well. While more realistic confining potentials such as a finite well would result in perturbations to the electronic structure, the key features are well described within our approximate model. This leads to the following envelope function 
\begin{equation}
F_{n,i} = \sqrt{\frac{8 V_c}{w^2L}}\sin\left ( \frac{n_x \pi x}{w}\right )\sin\left ( \frac{n_y \pi x}{w}\right )\sin\left ( \frac{n_z \pi z}{L}\right ),
\end{equation}
where $n=(n_x,n_y,n_z)$ and $V_c$ is the volume of the diamond unit cell.

The true eigenstates of the nanowire conduction band are linear combinations of the valley wavefunctions~\ref{wav} which respect its $\text{D}_{4\text{h}}$ symmetry. As presented in supplementary section~A, symmetry analysis reveals four degenerate conduction band minima states formed from the valleys with $\vec{K}_i$ perpendicular to the wire axis (with symmetries $\text{A}_{1g}$, $\text{B}_{1g}$ and $\text{E}_{u}$). Similarly, the valley states with $\vec{K}_i$ parallel to the wire axis form degenerate eigenstates with $\text{A}_{1g}$ and $\text{A}_{2u}$ symmetry. For $L\gg w$, the energy of the two-fold degenerate states exceeds that of the four-fold degenerate states and so only the latter are relevant for STIRAP. For the simple case of a wire aligned along a (100) axis, the eigenenergies of the four-fold conduction band minima states are given by
\begin{equation}\label{enWire}
E_n = \frac{\hbar^2\pi^2}{2}\left (\frac{n_x^2}{m_\parallel w^2} +\frac{n_y^2}{m_\perp w^2}+\frac{n_z^2}{m_\perp L^2}\right ),
\end{equation}
where $m_\perp$ and $m_\parallel$ are the transverse and longitudinal effective masses of an electron in bulk diamond.
For $L\gg w$, this corresponds to
\begin{equation}\label{delF}
\Delta E_c = \frac{3 \hbar^2\pi^2}{2m_\perp L^2}.
\end{equation}
In Section~\ref{sec:optimization} we use Equation~\ref{delF} to assess the feasibility condition $\Delta E_c/\hbar\gg\Omega$ for the two different nanowire designs. 

EMT is limited as it cannot account for coupling between valley states via the wire confining potential. Known as the valley-orbit interaction, this coupling leads to a fine splitting of the conduction band minima states\cite{Murphy-Armando2010}. While the resulting eigenstates may be determined by symmetry, the magnitude of the splitting is notoriously difficult to quantify but assumed to be small\cite{Resca1979,Resta1977,Luttinger1955,Altarelli1977}. A proper treatment would require detailed \textit{ab-initio} calculations, which are left as an opportunity for future work. The final electronic consideration is splitting induced by the applied electric field. As explored in supplementary section~A, this Stark effect does not influence states of the conduction band minimum. While it does produce a small splitting of the $\text{A}_{1g}$ and $\text{A}_{2u}$ states, for $L\gg w$ this poses no impediment to the performance of STIRAP.

\section{Optical processes in diamond nanowires}
\label{sec:optical}

Assessing the feasibility of STIRAP requires comparing attainable photoionization Rabi frequencies to optical decoherence mechanisms. In this section we first present calculations of the photoionization Rabi frequency including Franck-Condon effects. We then identify and characterize three optical decoherence mechanisms; erroneous capture by $\text{N}_\text{S}^+$ defects, e-p scattering and spontaneous emission.

\subsection{Photoionization Rabi frequency}

The conduction dipole moment was first obtained for the photoionization of bulk NV centers using \textit{ab-initio} calculations. Density functional theory was performed using the VASP plane-wave code with a 512-atom diamond supercell\cite{Kresse1996a,Kresse1994a,Kresse1996,Joubert1999} and Heyd-Scuseria-Ernzerhof functionals\cite{Heyd2003a}. The energy of the ${}^3\text{A}_2$ ground state of the $\text{NV}^{-}$ center and the ${}^2\text{E}$ ground state of the $\text{NV}^{0}$ center (plus an ionized electron) were calculated as a function of generalized atomic coordinates. As detailed in supplementary section~C, the calculations produced an ionization energy of $\omega=2.6$ eV (close to experimental values\cite{Aslam2013}) and a Huang-Rhys factor of $S=1.39$. Direct calculation of the optical matrix elements yields a transition dipole moment of $d_\text{bulk}=0.085 \ e \cdot \text{\AA}$, normalised to the volume of the 512-atom supercell.

The bulk photoionization dipole moment can be used to determine the corresponding moment in a nanowire. Noting that this moment is identical for both the pump and Stokes pulse, we calculate in supplementary section~C that
\begin{equation}\label{DMwire}
d_\text{wire} = F_0(\vec{R}_A) e^{-S/2} d_\text{bulk},
\end{equation}
where $\vec{R}_A$ is the position of NV $A$. For defects centered at opposite ends of the wire, the normalization function in Equation~\eqref{DMwire} can be simplified as
\begin{equation}\label{norm}
F_0(s) = \sqrt{\frac{8 V_\text{sc}}{w^2 L}}\cos\left ( \frac{\pi s}{2 L} \right ),
\end{equation}
where $V_\text{sc}=2.837\text{ nm}^3$ is the volume of the 512-atom supercell. Assuming a laser power $P$ with radial spot-size $r$, the effective photoionization Rabi frequency within a diamond nanowire is
\begin{equation}\label{RFbaby}
\Omega = \frac{e^{-S/2}}{r\hbar}\sqrt{\frac{8P}{n_Dc\epsilon_0\pi}}F_0(s)d_\text{bulk},
\end{equation}
where $n_D$ is the index of refraction in diamond. The achievable spot size and power are related to each other depending on the technical aspects of the optics employed. A conservative estimate says that a diffraction limited spot size $r=200$~nm can be attained with a lasing power of $P=100$~mW. These parameters will be used in Section~\ref{sec:optimization} to evaluate the photionization Rabi frequency.

\subsection{Decoherence mechanisms}

One source of optical decoherence is erroneous capture of conduction band electrons by $\text{N}_\text{S}^+$ defects. Typically $\text{N}_\text{S}$ defects are formed during CVD growth due to N-based gases present in the plasma\cite{Samlenski1995}. These defects act as electron donors, readily ionizing to $\text{N}_\text{S}^+$ to fill electron traps from surface defects\cite{Stacey2019}. The $\text{N}_\text{S}^+$ capture rate can be estimated as $\Gamma_\text{cap}=\rho_{\text{N}_\text{S}^+}\rho_\text{e}L^2w\sigma\sqrt{k_B T/m^*}$, where $\sigma\approx3-7 \ \text{nm}^2$ is the $\text{N}_\text{S}^+$ capture cross section, $k_B$ is Boltzmann's constant, $T$ is temperature, $m^*$ is the isotropic mean of the effective mass components, $\rho_{\text{N}_\text{S}^+}$ is the density of $\text{N}_\text{S}^+$ defects and $\rho_\text{e}$ is the probability density of the conduction band states\cite{Doherty2016}. We may conservatively estimate that $\rho_\text{e}L^2w<1$ and thus the capture rate is independent of volume. Assuming $T=4$ K and $\rho_{\text{N}_\text{S}^+}\approx 1$~ppb we find that $\Gamma_\text{cap}\lesssim 2.5$ MHz. This places a constraint on the photoionization Rabi frequency for STIRAP in any nanostructure. In this analysis we have assumed passivation of all surface traps by $\text{N}_\text{S}$ donors. If this is not the case surface defects must be considered as another source of electron capture.

A second source of decoherence can be attributed to e-p scattering of the conduction band states. We consider only first order processes involving adsorption and emission of acoustic phonons with higher energy conduction states. Higher order interactions can be considered negligible at $4$ K. The e-p scattering rate, $\Gamma_\text{ep}$, may be determined through Fermi's golden rule with the Hamiltonian 
\begin{equation}\label{HAMep}
\hat{\mathcal{H}}_\text{ep} = \Xi_d \vec{\nabla}\cdot \vec{u}_f,
\end{equation}
where $\Xi_d=8.7$ eV is the diamond deformation potential\cite{Prelas1997} and $\vec{u}_f$ is the nanowire phonon field. Explicit rates for $\Gamma_\text{ep}$ are specific to the nanowire design and are presented in the following section.

A fundamental source of optical decoherence is that due to spontaneous emission. The spontaneous emission rate for transitions $|2\rangle \rightarrow |1\rangle$ and $|2\rangle\rightarrow|3\rangle$ may be estimated using conventional expressions as\cite{Fox2006}
\begin{equation}\label{SE}
\Gamma_\text{SE}=2\frac{\omega^3 d_\text{wire}^2}{3\pi\epsilon_0 n \hbar c^3}.
\end{equation}
Note that this decoherence rate is dependent on wire geometry through the normalization function $F$ as per Equation~\eqref{DMwire}. We now evaluate the Rabi frequency and total decoherence rate within both nanowire designs -- surface confinement and electrostatic confinement -- and optimize wire dimensions to satisfy the STIRAP feasibility conditions.

\section{Optimization of nanowire design}
\label{sec:optimization}

\subsection{Surface Confinement}

As displayed in Figure~\ref{fig:wires}~\textbf{(A)}, electronic confinement may be realized through fabrication of free-standing nanowires\cite{Babinec2010,Burek2012} or nanopillars\cite{Momenzadeh2015} using reactive ion etching. To avoid injection of spurious electrons, it is necessary that the electrodes affixed to the wire ends are insulating and separated from the diamond surface with a high resistivity contact. NV centers may be introduced into wires through either N ion-implantation\cite{Bayn2015,Pezzagna2010} or N $\delta$-doping\cite{Ohashi2013}. A pair of NV centers suitable for STIRAP can then be identified through confocal microscopy. Ideally, the NVs will be co-aligned along a (111) axis and situated at depths greater than $\sim50$~nm (this is necessary for obtaining optimal bulk-like properties\cite{Ohno2012}).

We now derive an expression for the decoherence rate due to e-p scattering. In supplementary section~D, we present calculations of $\Gamma_\text{ep}$ by approximating $\vec{u}_f$ through elasticity theory. Acoustic nanowire phonon modes may be classified as dilational, flexural, torsional and shear\cite{Nishiguchi1997}. However, only dilational modes possess non-zero divergence and contribute to the e-p scattering as per Hamiltonian~\eqref{HAMep}. The dilational modes $\vec{u}_{d,m}$ have angular frequencies quantized by $m=(m_x,m_y,m_z)$ which are given by
\begin{equation}
\omega_m^2 = \pi^2 c_l^2\left ( \frac{m_x^2+m_y^2}{w^2} + \frac{m_z^2}{L^2} \right ),
\end{equation}
where $c_l$ is the longitudinal speed of sound in diamond. The scattering rate can then be calculated as
\begin{align}
\Gamma_\text{ep}=\frac{2\pi}{\hbar^2}\sum_{n,m}&\Xi_d^2|M_{n,m}|^2\frac{\hbar}{2\rho_C w^2L \omega_m} \nonumber \\
&\times n_B(\omega_m,T)\rho(\omega_n-\omega_m),\label{G_ns}
\end{align}
where $M_{n,m}$ is the overlap integral between the electron and phonon wavefunctions, $\rho_C$ is the density of diamond and $n_B$ is the Bose-Einstein distribution. The density of states, $\rho$, is assumed to be a Lorentzian, given by
\begin{equation}\label{lorentz}
\rho(\omega_n-\omega_m)=\frac{\gamma/\pi}{(\omega_n-\omega_m)^2+\gamma^2},
\end{equation}
where $\gamma\approx \omega_m/Q$ for $Q$ the quality factor. For diamond microcantilevers, $Q>10^6$ has been observed\cite{Tao2014}.

\begin{figure}
    \centering
    \begin{subfigure}[b]{0.5\textwidth}
        \centering\includegraphics[width=0.7\textwidth]{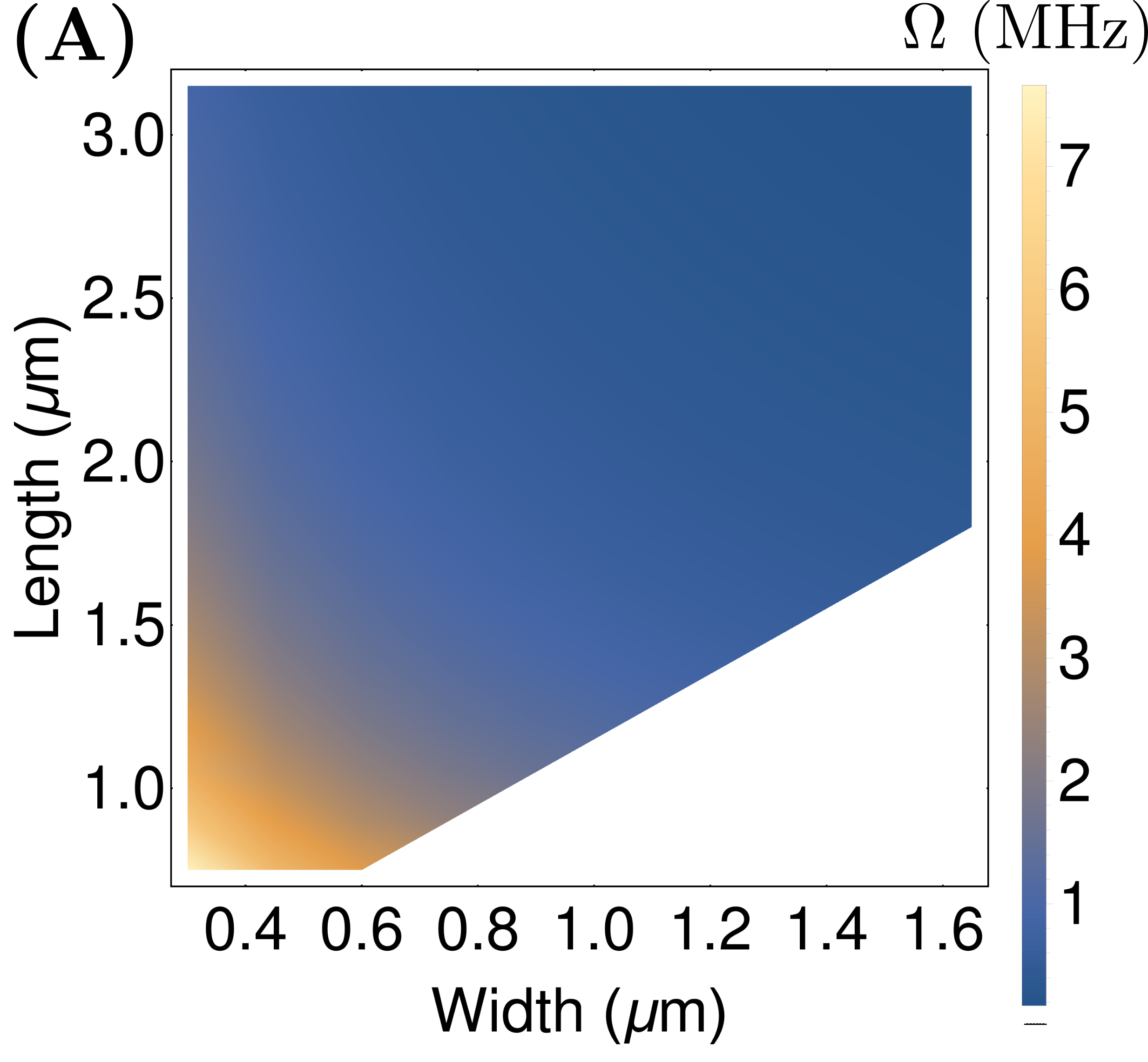}
    \end{subfigure}
    
    \begin{subfigure}[b]{0.5\textwidth}
        \centering\includegraphics[width=0.7\textwidth]{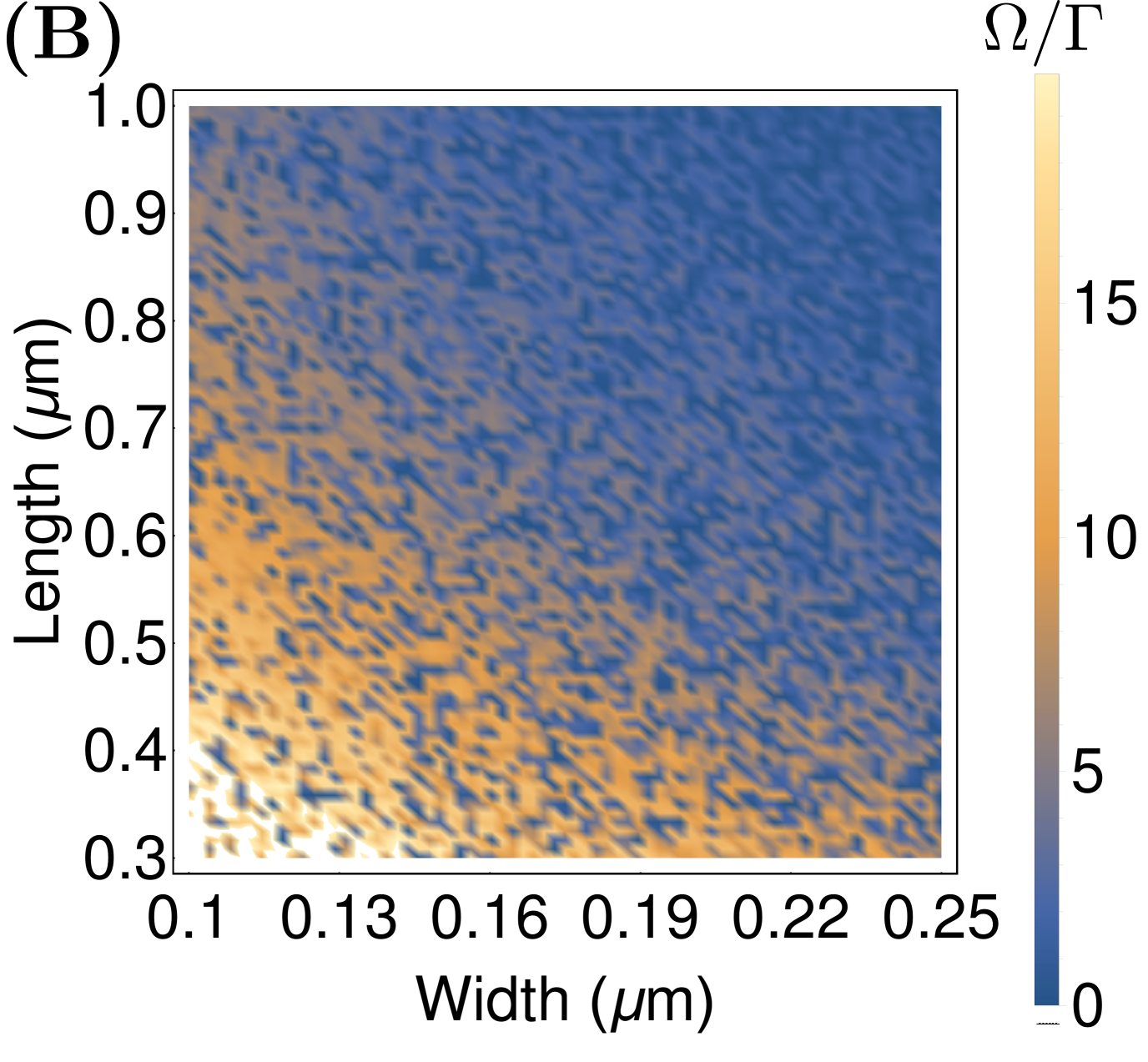}
    \end{subfigure}

    \begin{subfigure}[b]{0.5\textwidth}
       \centering\includegraphics[width=0.7\textwidth]{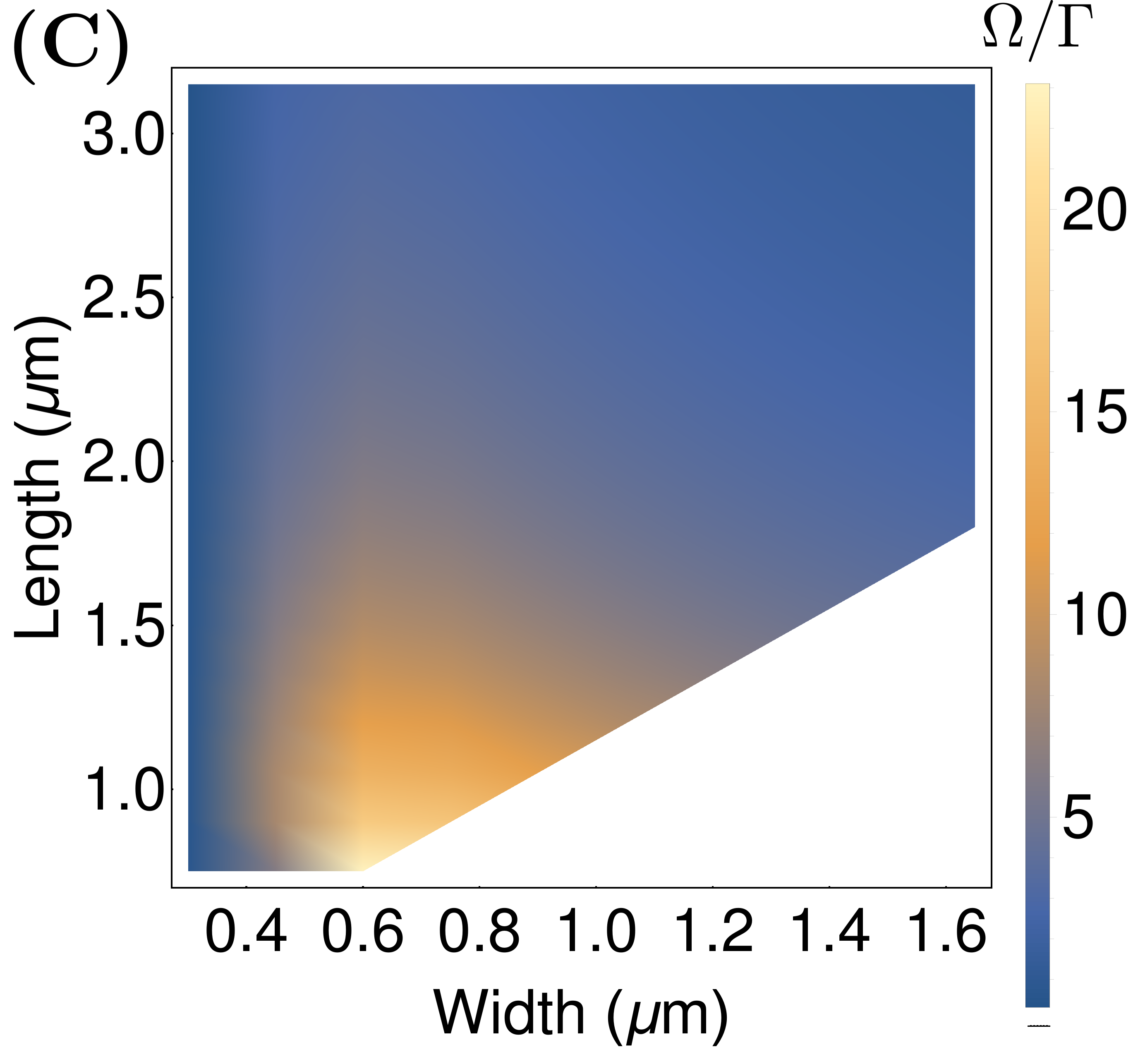}
       \label{fig:epNE}
    \end{subfigure}
    \caption{Dimensional dependence of optical properties within (100) aligned diamond nanowires.
        \textbf{(A)} Density plot of the photionization frequency $\Omega$ as a function of wire dimension.
        \textbf{(B)} Density plot of the ratio of the photoionization frequency to the optical decoherence rate ($\Omega/\Gamma$) as a function of dimension for a surface confined nanowire.
        \textbf{(C)} Density plot of the ratio of the photoionization frequency to the optical decoherence rate ($\Omega/\Gamma$) as a function of nanowire dimension for an electrostatically confined nanowire.
        Note we have omitted calculations of $\Omega$ and $\Omega/\Gamma$ for wires with $w>L$.}\label{fig:densities}
\end{figure}

We now optimize the dimensions of a surface confined nanowire with respect to the STIRAP feasibility condition $\Delta E_c /\hbar\gg\Omega\gg\Gamma$. In Figure~\ref{fig:densities}~\textbf{(A)} we present the photionization Rabi frequency derived in Equation~\eqref{RFbaby} as a function of wire dimension. In Figure~\ref{fig:densities}~\textbf{(B)} we compare the Rabi frequencies to the total decoherence rate $\Gamma=\Gamma_\text{cap}+\Gamma_\text{SE}+\Gamma_\text{ep}$ for varying nanowire dimension. We assume that $Q=10^6$, $T=4 \ \text{K}$ and that each NV center is positioned 100~nm from its respective wire end. For simplicity we have only considered wires aligned along a (100) axis, though similar results likely hold for other wire directions.

Figure~\ref{fig:densities}~\textbf{(B)} indicates that in general decoherence effects are minimized for smaller wire dimensions. This can be attributed to the normalization of the Rabi frequency, which scales inversely proportional with wire volume as evident in Figure~\ref{fig:densities}~\textbf{(A)} (c.f., Equation~\ref{RFbaby}). The sporadic variations in $\Omega/\Gamma$ can be attributed to resonance between electronic and phononic levels. Consulting equation~\eqref{G_ns}, certain wire dimensions produce an increased density of resonant states which amplifies the rate of e-p scattering. Off resonance, the dominant decoherence mechanism is electron capture. A range of wire dimensions with $L<0.6$~$\mu\text{m}$ and $w<0.2$~$\mu\text{m}$ satisfy the STIRAP feasibility condition with at least $\Omega>10\Gamma$, reaching a maximum of $\Omega\approx20\Gamma$ for short wires 0.3~$\mu\text{m}$ long. Note that for all wire dimensions considered in Figure~\ref{fig:densities}~\textbf{(B)}, $\Delta E_c/\hbar>10^3 \Omega$, and therefore off-resonant coupling does not pose an impediment to STIRAP.

\subsection{Electrostatic confinement}

An alternative means for realizing a nanowire confining potential is through electrodes in a bulk structure. Consider the diamond substrate presented in Figure~\ref{fig:wires}~\textbf{(B)} where two co-aligned NV centers can be identified through confocal microscopy. To generate the confining potential, a square nano-electrode with width $w$ is affixed to the surface directly above the NV centers. On the opposing surface of the diamond substrate an electrode plate is fixed and grounded. Applying a positive potential to the square electrode will then generate a confining potential. This localizes the electron density near the surface into a wire-like geometry. The effective length of this wire may be tuned through the magnitude of the potential. Conveniently, the potential also generates an electric field gradient which can be used to energetically distinguish each NV center during STIRAP. These electrostatic properties are demonstrated through simulations using COMSOL Multiphysics\textregistered \ software presented in supplementary section~E.

In contrast to surface confined wires, electron scattering within electrostatically confined wires is due to interactions with bulk phonons. Calculation of $\Gamma_\text{ep}$ using Fermi's golden rule and the Hamiltonian~\eqref{HAMep} is presented in supplementary section~D. We find that
\begin{align}
\Gamma_\text{ep}=&\frac{1}{2(2\pi)^2}\frac{\Xi_d^2}{\hbar\rho_C c_l^7}\sum_n \omega_n^5 n_B(\omega_n,T) \nonumber\\
&\times\int_0^\pi\int_0^{2\pi}G_n\left ( \frac{\omega_n}{c_l},\theta,\phi\right )\sin(\theta)\text{d}\theta\text{d}\phi,\label{G_ne}
\end{align}
where $G_n$ is a complicated expression for the overlap integral.

A further benefit of the electrostatic confinement design is that decoherence due to $\text{N}_\text{S}^+$ capture can be mitigated through inclusion of a sacrificial donor layer. $\text{N}_\text{S}$ defects distant from the confining potential can be $\delta$-doped into the substrate during CVD growth\cite{Ohashi2013}. The sacrificial defects donate electrons to the surface traps until charge conservation occurs, reducing the density of $\text{N}_\text{S}^+$ available for erroneous capture within the confining potential\cite{Stacey2019}. We conservatively estimate that the presence of a sacrificial layer can reduce $\Gamma_\text{cap}$ by 95\%. 

In Figure~\ref{fig:densities}~\textbf{(C)} we present the dimensional dependence of $\Omega/\Gamma$ for an electrostatically confined nanowire aligned along the (100) axis. We have that $G_n\propto w^{-4}$ in Equation~\ref{G_ne} and therefore e-p scattering dominates the Rabi frequency for wire widths less than approximately $0.4 \ \mu\text{m}$. For larger wire dimensions, both e-p scattering and spontaneous emission are negligible and the dominant source of decoherence is due to $\text{N}^+_\text{S}$ capture. The STIRAP feasibility conditions are satisfied when $0.5 \ \mu\text{m}<w< 0.9 \ \mu\text{m}$ and $0.75 \ \mu\text{m}<L<1.25  \ \mu\text{m}$ which yields $\Omega>10\Gamma$. The ratio $\Omega/\Gamma$ is optimized at $\approx 25$ for wire dimensions of $L=0.75 \ \mu\text{m}$ and $w = 0.6 \ \mu\text{m}$. Furthermore, for all wire dimensions $\Delta E_c/\hbar\gg10^3\Omega$ and hence cross-talk with higher energy states in the conduction band is negligible. 

\section*{Conclusion}
\label{sec:conc}
In this paper we propose spatial STIRAP to realize an on-chip spin quantum bus for scalable diamond QIP devices. Our scheme considers coherent quantum transport of an electron and its spin-state between two NV centers embedded in a diamond nanowire. This is achieved through implementing STIRAP for optical control of the NV center charge states and the confined nanowire conduction states. In contrast to existing spatial STIRAP protocols which manipulate tunneling amplitudes between distant sites, our proposal realizes quantum transport without any occupation of the intermediary space. Furthermore, to our knowledge this is the first implementation of spatial STIRAP including spin-state transport.

A proposal for our scheme was first presented along with a protocol for initializing the NV spin-states. We identified a feasibility condition for implementing STIRAP, $\Delta E_c/\hbar\gg\Omega\gg\Gamma$, requiring the photionization Rabi frequency to exceed optical decoherence rates and be limited by coupling with higher-energy conduction states. For the latter, EMT was employed to model the effects of the nanowire confining potential on the diamond conduction band. We then presented the first \textit{ab-initio} calculations of the $\text{NV}$ photoionization Rabi frequency. This allowed the STIRAP feasibility condition to be evaluated in two nanowire designs, which achieved electronic confinement through either a surface structure or electrodes. By optimizing the Rabi frequency relative to the decoherence rate for varying wire dimension, we conclude that STIRAP is feasible in both designs in timescales on the order of hundreds of nanoseconds.

The first experimental stages should perform photionization spectroscopy to assess the spectral line width of discretized states in the nanowire conduction band. Ultimately, this will determine whether deterministic photoionization to the conduction band minimum is feasible. This study has also introduced multiple avenues for future theoretical research. For example, a detailed \textit{ab-initio} model of the nanowire electronic structure would allow for precise calculation of valley-orbit and spin-orbit interactions. This would require development of new computational tools for simulating mesoscopic diamond structures.

\section*{Acknowledgements} 
We acknowledge funding from the Australian Research Council (DE170100169, DP170103098, FT160100357, CE170100026). C.A.M acknowledges support from the National Science Foundation through grants NSF-1619896, NSF-1547830, and from Research Corporation for Science Advancement through a FRED Award. AA acknowledges funding from the European Union's Horizon 2020 research and innovation programme under grant agreement No. 820394 (project Asteriqs). This research was undertaken with the assistance of resources and services from the National Computational Infrastructure (NCI), which is supported by the Australian Government.

\bibliographystyle{spiejour}   
\bibliography{STIRAP}   

\end{document}


\title{\vspace{-4.5cm}Supplementary: Spin coherent quantum transport of electrons between defects in diamond}
\author[a]{L. M. Oberg}
\author[a]{E. Huang}
\author[a]{P. M. Reddy}
\author[b]{A. Alkauskas}
\author[c]{A. D. Greentree}
\author[d]{J. H. Cole}
\author[a]{N. B. Manson}
\author[e]{C. A. Meriles}
\author[a]{M. W. Doherty}
\affil[a]{Laser Physics Center, Research School of Physics and Engineering, Australian National University, Australian Capital Territory 2601, Australia}
\affil[b]{Center for Physical Sciences and Technology, Vilnius LT-10257, Lithuania}
\affil[c]{ARC Center of Excellence for Nanoscale BioPhotonics, School of Science, RMIT University, Melbourne, VIC, 3001, Australia}
\affil[d]{Chemical and Quantum Physics, School of Science, RMIT University, Melbourne, Victoria 3001, Australia}
\affil[e]{Department of Physics, CUNY-City College of New York, New York, New York 10031, USA}

\maketitle

\section*{Section A: Electronic structure of diamond nanowires}

\subsection*{Effective mass theory}
Approximate analytical forms for the electronic states of a diamond nanowire can be obtained using effective mass theory (EMT)\cite{Stoneham1976}. Consider the diamond crystal Hamiltonian $\hat{\mathcal{H}}_0$ which has been solved as
\begin{equation}
\hat{\mathcal{H}}_0 \phi_{n,\vec{k}}=\mathcal{E}_n(\vec{k})\phi_{n,\vec{k}},
\end{equation}
for the energies of the $n^\text{th}$ conduction band $\mathcal{E}_n(\vec{k})$ corresponding to the Bloch wavefunctions $\phi_{n,\vec{k}}=e^{i \vec{k}\cdot \vec{r}}u_{n,\vec{k}}(\vec{r})$, where $u_{n,\vec{k}}(\vec{r})$ is some lattice periodic function. We desire to solve the Schr\"odinger equation
\begin{equation}\label{val}[\hat{\mathcal{H}}_0 + U(\vec{r})]\Psi(\vec{r})=E\Psi(\vec{r}),\end{equation}
where $U(\vec{r})$ is the confining potential of the nanowire (aligned along the $\hat{z}$ axis) given by
\begin{equation}\label{pot}
U(x,y,z)=
\begin{cases}
0, \ \ \ \text{if $0< x,y <w$ and $0<z<L$;}\\
\infty, \ \ \ \text{otherwise.}\\
\end{cases}
\end{equation}
To do so, we expand the wavefunction $\Psi(\vec{r})$ into the basis of Bloch states,
$$\Psi(\vec{r}) = \sum_{n,\vec{k}} F_n(\vec{k})\phi_{n,\vec{k}}(\vec{r}),$$
where $F_n(\vec{k})$ is the Fourier transform of some envelope function $F_n(\vec{r})$. $E$ and $F_n(\vec{r})$ can be solved at the nanowire conduction band minimum through three assumptions:
\begin{enumerate} \item $U(\vec{r})$ is slowly varying with respect to the lattice spacing.
\item $F_n(\vec{k})$ is only significant well within the first Brillouin zone (that is, for $|\vec{k}| \ll \frac{\pi}{a}$).
\item The dispersion relationship near the conduction band minimum (termed a valley) can be approximated using the effective mass tensor, $\overleftrightarrow{m^*}$, of the bulk crystal
\begin{equation}
\label{emt} E_c(\vec{k}) \approx E_c + \frac{\hbar^2}{2}\vec{k}\cdot \overleftrightarrow{m^*}\cdot \vec{k},\end{equation}
where $E_c$ is the energy of the conduction band minimum.
\end{enumerate}
By satisfying these requirements a position space representation for $F_n(\vec{k})$ near the conduction band minimum can be obtained by solving
\begin{equation}\label{1d}
\Bigg [ -\frac{\hbar^2}{2}\vec{\nabla}\cdot \overleftrightarrow{m^*} \cdot\vec{\nabla} + E_c + U(\vec{r}) \Bigg ]F_n(\vec{r}) = E_n F_n(\vec{r}).
\end{equation}

\begin{figure}[]
	\centering
	\includegraphics[width=0.7\textwidth]{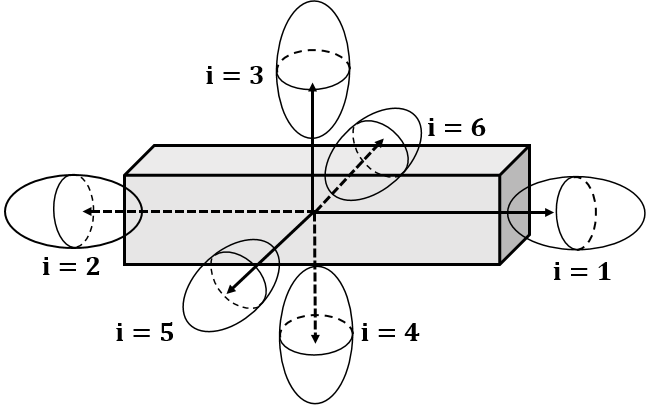}
	\caption{The six equivalent conduction band minima states of bulk diamond are projected on to the nanowire $k$-space. This is visualized through six dispersion relations, dubbed valleys, represented by ellipsoidal energy iso-contours due to diamond's anisotropic effective mass tensor. For a wire with its axis parallel to the (100) direction, a single iso-contour is positioned at each face of the rectangular prism.}
	\label{fig:valleys}
\end{figure}

Bulk diamond possesses an fcc reciprocal lattice with six equivalent valleys located between the $\Gamma$ and $X$ points. For a nanowire aligned along the (100) axis, these valleys are projected onto each face of the rectangular prism and may be enumerated as $i=1,\ldots,6$. This is depicted in Figure~\ref{fig:valleys}, where the valleys are represented by ellipsoidal energy iso-contours in k-space due to diamond's anisotropic mass tensor. The envelope function for each valley can be solved using Equation~\eqref{1d}, yielding eigenstate solutions for Equation~\ref{val} given by
\begin{equation}\label{sols}
\psi_{n,i}(\vec{r}) = F_{n,i}(\vec{r})e^{i\vec{K_i}\cdot \vec{r}}u_{0,i}(\vec{r}),
\end{equation}
where $\vec{K_i}$ is the bulk reciprocal vector for the valley in question.

Explicitly solving Equation~\eqref{1d} for the nanowire confining potential yields the envelope function
\begin{equation}
F_{n,i} = \sqrt{\frac{8 V_c}{w^2L}}\sin\left ( \frac{n_x \pi x}{w}\right )\sin\left ( \frac{n_y \pi x}{w}\right )\sin\left ( \frac{n_z \pi z}{L}\right ),
\end{equation}
where $n=(n_x,n_y,n_z)$, $V_c$ is the volume of the diamond unit cell, and we have used the normalization condition $\langle \psi_{n,i}|\psi_{n,j}\rangle=\delta_{ij}$. The corresponding eigenenergies for a wire along the (100) axis are 
\begin{equation}\label{ens12}
E_n^{i=1,2} = \frac{\hbar^2\pi^2}{2}\left (\frac{n_x^2}{m_\perp w^2} +\frac{n_y^2}{m_\perp w^2}+\frac{n_z^2}{m_\parallel L^2}\right ),
\end{equation}
for valleys 1 and 2,
\begin{equation}\label{ens34}
E_n^{i=3,4} = \frac{\hbar^2\pi^2}{2}\left (\frac{n_x^2}{m_\parallel w^2} +\frac{n_y^2}{m_\perp w^2}+\frac{n_z^2}{m_\perp L^2}\right ),
\end{equation}
for valleys 3 and 4, and
\begin{equation}\label{ens56}
E_n^{i=5,6} = \frac{\hbar^2\pi^2}{2}\left (\frac{n_x^2}{m_\perp w^2} +\frac{n_y^2}{m_\parallel w^2}+\frac{n_z^2}{m_\perp L^2}\right ),
\end{equation}
for valleys 5 and 6. Solutions for wires aligned along other axes can be obtained through rotations of the effective mass tensor.

The true eigenstates of the nanowire must be commensurate with the $D_{4h}$ symmetry of the wire. A symmetry respecting basis can be constructed through linear combinations of the single valley EMT solutions presented in Equation~\eqref{sols} which agree with the irreducible representations of $D_{4h}$\cite{Dresselhaus2007}. Standard application of the projection operators yields the following. For the eigenspace spanned by $\{\psi_{n,1}, \psi_{n,2}\}$, the symmetry respecting basis for the $n=(1,1,1)$ conduction band minima states is
\begin{align*}
|A_{1g,1}\rangle&=\frac{1}{\sqrt{2}}(\psi_{n,1} + \psi_{n,2}), \\
|A_{2u}\rangle&=\frac{1}{\sqrt{2}}(-\psi_{n,1} + \psi_{n,2}),
\end{align*}
while for the eigenspace spanned by $\{\psi_{n,i}\}_{i=3}^6$, 
\begin{align*}
|A_{1g,2}\rangle&=\frac{1}{2}(\psi_{n,3} + \psi_{n,4}+\psi_{n,5} + \psi_{n,6}), \\
|B_{1g}\rangle&=\frac{1}{2}(-\psi_{n,3} - \psi_{n,4}+\psi_{n,5} + \psi_{n,6}),\\
|E_{u,1}\rangle&=\frac{1}{\sqrt{2}}(-\psi_{n,5} + \psi_{n,6}),\\
|E_{u,2}\rangle&=\frac{1}{\sqrt{2}}(-\psi_{n,3} + \psi_{n,4}).
\end{align*}
For $L>>w$, the two-fold degenerate states $|A_{1g}\rangle$ and $|A_{2u}\rangle$ are significantly higher in energy than the four-fold degenerate states.

\subsection*{Stark effect}

Implementation of STIRAP requires energetically distinguishing the two NV centers in the nanowire. This is achieved through application of a potential difference $\Phi$ between each end of the length $L$ wire. The corresponding Hamiltonian is given by
\begin{equation}\label{Hstark}
\hat{\mathcal{H}}_\text{Stark}=\frac{e\Phi}{L}z.
\end{equation}
Note that the Hamiltonian~\eqref{Hstark} transforms as $z$ and therefore has $A_{2u}$ symmetry. Consulting the $D_{4h}$ product table\cite{Dresselhaus2007}, non-zero coupling only occurs between the conduction band eigenstates with $A_{2u}$ and $A_{1g}$ symmetry. For $L>w$, these are not the states of the conduction band minimum. Further calculations of the perturbing matrix elements reveal that the Stark effect induces only kHz splitting of the $|A_{2u}\rangle$ and $|A_{1g,1}\rangle$ states for wires with $\sim\mu$m dimensions and applied voltages of approximately 1 V. Hence, we conclude that the Stark effect does not present any impediment to the implementation of STIRAP.

\section*{Section B: Spin-orbit coupling in diamond nanowires}

\begin{figure}[]
	\centering
	\includegraphics[width=1.0\textwidth]{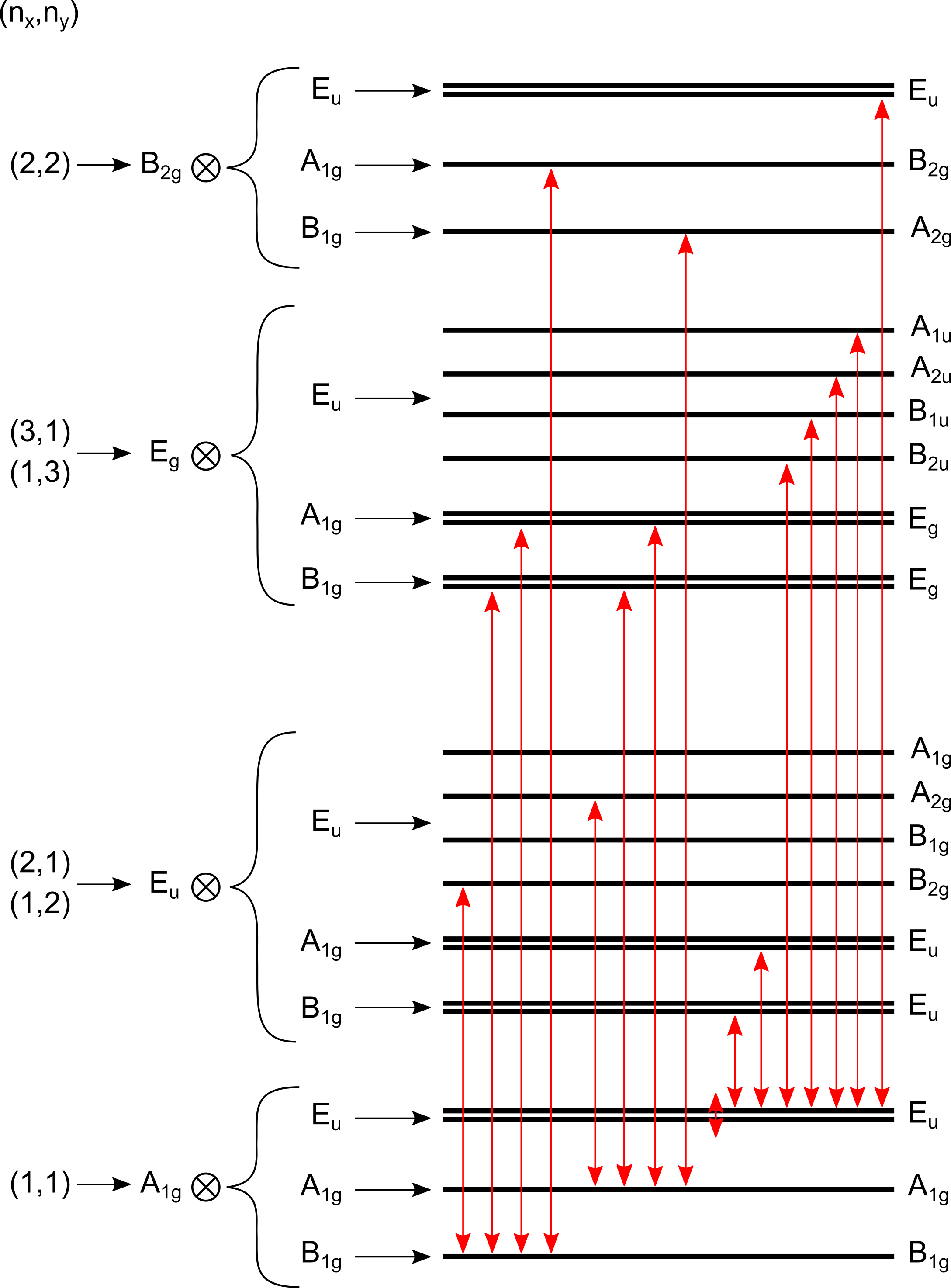}
	\caption{The allowed second-order spin-orbit couplings between the four-fold degenerate conduction band minima states and their excitations. The large number of possible couplings results in a severely complicated system, requiring extensive \textit{ab-initio} data to completely model. As a result, we resort to the closure approximation to derive an effective model for spin-orbit coupling.}
	\label{fig:SOcoupling}
\end{figure}

In this section we investigate spin-orbit interactions within the nanowire conduction band minima states. Consider a nanowire confining potential aligned along the $\hat{z}$ axis as defined in Equation~\eqref{pot}. The spin-orbit Hamiltonian reads as
\begin{equation}\label{SO}
\hat{\mathcal{H}}_\text{SO}=\alpha (\vec{\nabla }V \times \vec{p})\cdot \vec{s},
\end{equation}
where $\alpha$ is the coupling constant and $V=V_b+U$ is the full nanowire potential; the sum of the bulk electrostatic potential experienced by the conduction electrons, $V_b$, and the wire confining potential, $U$. Consider the symmetries of the angular momentum components of $\hat{\mathcal{H}}_\text{SO}$. In $D_{4h}$, $(\hat{L}_x,\hat{L}_y)$ transforms as $E_g$ and $\hat{L}_z$ as $A_{2g}$. By the Wigner-Eckert theorem\cite{Dresselhaus2007}, these symmetries allow for many possible first and second-order couplings between the lowest energy valley states and their excitations. A level scheme of these couplings is visualized in Figure~\ref{fig:SOcoupling}. Note that we assume that $L>>w$ and therefore do not consider the $|A_{1g,1}\rangle$ and $|A_{2u}\rangle$ states and their excitations (as they are much higher in energy).

Figure~\ref{fig:SOcoupling} indicates that a full analytical model of spin-orbit coupling is highly complicated due to the sheer number of possible interactions. Not only this, but evaluation of any analytical expression would require extensive \textit{ab-initio} data for the nanowire conduction band. Consequently, as we currently lack such data, we will instead apply the closure approximation\cite{Sakurai1986} to produce an effective model of spin-orbit coupling to second order.

We shall treat the spin-orbit Hamiltonian as a perturbation to $\hat{\mathcal{H}}_0+U$ and work in the four-fold degenerate symmetry states of the nanowire conduction band minima. We therefore separate the $\hat{\mathcal{H}}_\text{SO}$ into its first and second-order contributions as
\begin{equation}\label{inte}
\hat{\mathcal{H}}_\text{SO}=\hat{\mathcal{H}}_\text{SO}^{(1)}+\hat{\mathcal{H}}_\text{SO}^{(2)}.
\end{equation}
Firstly, we note that by the Wigner-Eckart theorem\cite{Dresselhaus2007} the only non-zero first-order coupling occurs between the $E_u$ states via the $\hat{L}_z$ component of $\hat{\mathcal{H}}_\text{SO}^{(1)}$. That is,
$$\hat{\mathcal{H}}_\text{SO}^{(1)} = \lambda_\parallel^{(1)}\hat{L}_z \hat{s}_z.$$
This interaction lifts the degeneracy of the $E_u$ states, forming $E_{u-}$ and $E_{u+}$ states. These correspond with alignment and anti-alignment of the electron spin with its orbital angular momentum along the nanowire axis. This coupling, which we denote as having magnitude $\lambda_\parallel^{(1)}$, is well quantized in the $z$-axis. Hence, we write in the spin-orbit basis
\begin{equation}\label{HSO1}
\hat{\mathcal{H}}_\text{SO}^{(1)} = \lambda_\parallel^{(1)} (| E_{u+}\rangle \langle E_{u+}| - |E_{u-}\rangle \langle E_{u-} |),
\end{equation}
where
\begin{align*}\lambda_\parallel^{(1)} &= \langle E_{u+} | \hat{\mathcal{H}}_\text{SO}^{(1)} | E_{u+} \rangle =  -\langle E_{u-} | \hat{\mathcal{H}}_\text{SO}^{(1)} | E_{u-} \rangle.
\end{align*}

To second order, the closure approximation produces the perturbative term
\begin{equation}\label{here}
\hat{\mathcal{H}}_\text{SO}^{(2)}=\frac{1}{\Delta E}\vec{s}\cdot \overleftrightarrow{L}^{(2)} \cdot \vec{s},
\end{equation}
where $\Delta E$ is the average energy difference between the excited states and a given valley symmetry state, and $\overleftrightarrow{L}^{(2)}$ is the following matrix of operators
$$
\overleftrightarrow{L}^{(2)} = \begin{bmatrix}
\hat{L}_x^2 & \hat{L}_x\hat{L}_y & \hat{L}_x\hat{L}_z\\ 
\hat{L}_y\hat{L}_x & \hat{L}_y^2 &  \hat{L}_y\hat{L}_z \\
\hat{L}_z\hat{L}_x & \hat{L}_z\hat{L}_y & \hat{L}_z^2
\end{bmatrix}.
$$
Expanding~\eqref{here} out explicitly, we obtain 
\begin{align}
\hat{\mathcal{H}}_\text{SO}^{(2)}&=\frac{1}{\Delta E}\Big [ \hat{L}_z^2\hat{s}_z^2+\frac{1}{2}(\hat{L}_x^2+\hat{L}_y^2)(\hat{s}_x^2+\hat{s}_y^2) + \frac{1}{2}(\hat{L}_x^2-\hat{L}_y^2)(\hat{s}_x^2-\hat{s}_y^2) \nonumber \\
&+\frac{1}{2}(\hat{L}_x\hat{L}_y+\hat{L}_y\hat{L}_x)(\hat{s}_x\hat{s}_y+\hat{s}_y\hat{s}_x)+\frac{1}{2}(\hat{L}_x\hat{L}_y-\hat{L}_y\hat{L}_x)(\hat{s}_x\hat{s}_y-\hat{s}_y\hat{s}_x) \Big ] \nonumber\\
&= \frac{1}{\Delta E}\Big [ \frac{\hbar^2}{4}(\hat{L}_x^2+\hat{L}_y^2+\hat{L}_z^2)I_s + \frac{i\hbar}{2}(\hat{L}_x\hat{L}_y-\hat{L}_y\hat{L}_x)\hat{s}_z \Big ]\label{calipso},
\end{align}
where we have used the Pauli matrix anti-commutator relations and $I_s$ is the spin identity operator.

Equation~\eqref{calipso} has significant implications for spin-transport within nanowires as it indicates that spin is quantized purely along the $\hat{z}$-axis. If the NV centers are co-aligned with the longitudinal direction of the nanowire (i.e., if the nanowire is aligned along the (111) axis), then so are their spin-quantization axes. Hence, there does not exist any transverse spin-orbit interaction during transport with spatial STIRAP. Consequently, there exists no mechanism for e-p scattering to mix spin projections. Thus, to second order in the closure approximation, the nanowire confining potential does not induce any additional spin relaxation.

\section*{Section C: Evaluation of the photoionization Rabi frequency}

In this section we first derive an expression for the photoionization Rabi frequency of an NV center in a diamond nanowire. We identify an explicit dependence on the bulk NV photoionization dipole moment, which we then calculate using \textit{ab-initio} techniques.

\subsection*{Derivation of Rabi frequency}

Recall that the states of the STIRAP $\Lambda$ scheme are defined as
\begin{align*}
|1\rangle &= |\text{NV}^-\rangle_A|\text{NV}^0\rangle_B \\
|2\rangle &= |\text{NV}^0\rangle_A|\text{NV}^0\rangle_B|\psi\rangle \\
|3\rangle &= |\text{NV}^0\rangle_A|\text{NV}^-\rangle_B,
\end{align*}
where $|\psi\rangle$ is the state of the nanowire conduction band minimum. The pump and Stokes lasers will be modeled as classical fields within the dipole approximation. Hence, the Rabi frequency for the transition $|1\rangle\rightarrow|2\rangle$ (similarly $|2\rangle\rightarrow |3\rangle$) is given by\cite{Fox2006}
\begin{equation}\label{rabiFreq}
\Omega_{P}=\frac{\vec{d}_\text{wire}^{P}\cdot \vec{E}_0}{\hbar},
\end{equation}
where $\vec{d}_\text{wire}^{P}$ is the transition dipole moment for $|1\rangle\rightarrow|2\rangle$ and $E_0$ is the electric field of the laser pulse.

We may calculate the photoionization dipole moment including Franck-Condon effects as
\begin{align}
\vec{d}_\text{wire}^{P}&=\langle\mu_0(\vec{R}_A)|\nu_p(\vec{R}_A) \rangle\langle 1 | (-e\vec{r})|2\rangle \nonumber \\
&= \langle\mu_0(\vec{R}_A)|\nu_p(\vec{R}_A) \rangle \langle \text{NV}_A^{-}| (-e\vec{r})|\text{NV}^0_A\rangle|\psi\rangle \nonumber \\
&= -e\langle\mu_0(\vec{R}_A)|\nu_p(\vec{R}_A)\rangle \int_V {\text{NV}_A^{-}}^*(\vec{r}-\vec{R}_A)\vec{r}{\text{NV}_A^{0}}(\vec{r}-\vec{R}_A)F_0(\vec{r})u(\vec{r})\text{d}^3r, \label{d12_1}
\end{align}
where $|\mu_0\rangle$ is the vibrational ground state of $|\text{NV}^-\rangle_A$, $|\nu_p\rangle$ is the $p^\text{th}$ vibrational state of $|\text{NV}^0\rangle_A$, $\vec{R}_A$ is the position of NV $A$ and $V$ is the volume of the wire. Note that $|\text{NV}^{-}\rangle_A$ and $|\text{NV}_A^{0}\rangle$ are both highly localized wavefunctions around $\vec{R}_A$ and $u(\vec{r})$ varies quickly on the period of the unit cell. However, by construction $F_0(\vec{r})$ varies slowly over the period of a unit cell. Hence, we approximate expression~\eqref{d12_1} by summing over all unit cells $j$ in the nanowire as
\begin{align}
-\frac{\vec{d}_\text{wire}^{P}}{e\langle\mu_0(\vec{R}_A)|\nu_p(\vec{R}_A)\rangle}&\approx\sum_{j}F_0(\vec{R}_j)\int_\text{cell}{\text{NV}_A^{-}}^*(\vec{r}-\vec{R}_A)\vec{r}{\text{NV}_A^{0}}(\vec{r}-\vec{R}_A)u(\vec{r})\text{d}^3r \nonumber \\
&= F_0(\vec{R}_A)\int_\text{cell}{\text{NV}_A^{-}}^*(\vec{r}-\vec{R}_A)\vec{r}{\text{NV}_A^{0}}(\vec{r}-\vec{R}_A)u(\vec{r})\text{d}^3r,\label{d12_2}
\end{align}
for some position $\vec{R}_j$ within each unit cell $j$.

However, note that expression~\eqref{d12_2} is exactly
\begin{equation}\label{d12_bulk}
\vec{d}_\text{wire}^{P}=F_0(\vec{R}_A)\langle\mu_0|\nu_p\rangle\vec{d}_\text{bulk},
\end{equation}
where $\vec{d}_\text{bulk}$ is the photoionization dipole moment of the NV center in bulk diamond. The vibrational overlap integral can be estimated through Huang-Rhys theory, which states that
\begin{equation}\label{HuangRhys}
|\langle\mu_0|\nu_p\rangle|^2=e^{-S} \frac{S^p}{p!},
\end{equation}
where $S$ is the Huang-Rhys factor. For STIRAP we are only interested in resonant addressing and therefore $p=0$.

The amplitude of a classical electric field is given by\cite{Fox2006}
\begin{equation}\label{fieldA}
|\vec{E}_0|^2 = \frac{4P}{n_Dc\epsilon_0\pi r^2},
\end{equation}
for a laser power $P$, refractive index of diamond $n_D$ and Gaussian beam-width of radius $r$. Hence, inserting expressions~\eqref{d12_bulk} and~\eqref{fieldA} into equation~\eqref{rabiFreq}, the total Rabi frequency in the STIRAP framework is given by
\begin{equation}\label{NVRabi}
\Omega = \sqrt{\Omega_P^2 + \Omega_S^2}=\frac{e^{-S/2}}{r\hbar}\sqrt{\frac{8P}{n_Dc\epsilon_0\pi}}F_0(\vec{R}_A)d_\text{bulk}.
\end{equation}
Consequently, evaluation of the Rabi frequency requires a value for the photoionization dipole moment of the bulk NV center.

\subsection*{\textit{Ab-initio} calculations of the photoionization dipole moment for the bulk NV center}

Our first-principles calculations are based on the hybrid density functional of Heyd, Scuseria, and Ernzerhof\cite{Heyd2003a}. In this approach, a fraction $\alpha$ of screened Fock exchange is admixed to the short-range exchange potential described by the generalized gradient approximation of Perdew, Burke and Ernzerhof\cite{Perdew1996a}. We used the standard value $\alpha=0.25$, for which both the lattice constant $a=3.544$~\AA\ and the band gap $E_g=5.3$~eV are in very good agreement with experimental values. We used the projector-augmented wave approach\cite{Blochl1994a} with a plane-wave energy cutoff of 400~eV. Calculations have been performed using the program \textsc{vasp}\cite{Kresse1996a,Kresse1994a,Kresse1996,Joubert1999}. Calculations of the negatively charged NV center have been performed using a supercell containing 512 atomic sites ($4\times 4\times 4$ fcc cell, volume 2.837 nm$^3$). A sole $\Gamma$ point has been used to sample the Brillouin zone. Ionic relaxation was carried out until Hellman-Feynman forces were less than 0.005 eV/\AA. 

\begin{figure}[h]
	\centering
	\includegraphics[width=1.0\textwidth]{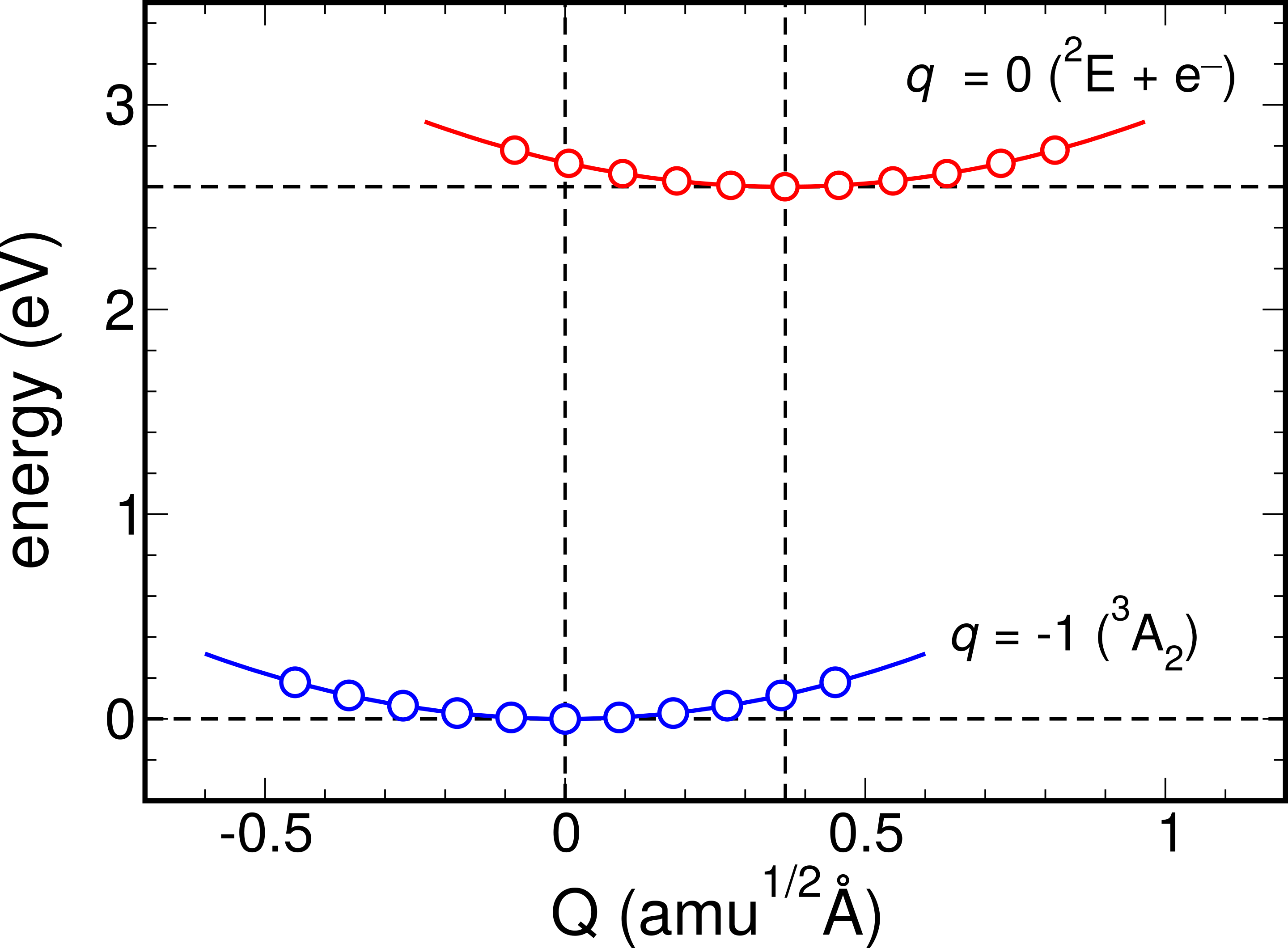}
	\caption{The energy of the ${}^3\text{A}_2$ ground state of the $\text{NV}^{-}$ center and the ${}^2\text{E}$ ground state of the $\text{NV}^{0}$ center (plus an ionized electron) as a function of generalized atomic coordinates. The \textit{ab-initio} calculations identified a Huang-Rhys factor of 1.39.}
	\label{fig:HRfactor}
\end{figure}

Momentum matrix element $p_{if}$ between $e$ states of the NV center and the conduction band minimum (CBM) has been calculated as follows. $p_{if}$ has been calculated for both occupied spin-majority $e$ states and each of the 6 CBM states (via the velocity matrix element formalism as implemented in \textsc{vasp}\cite{Kresse1996a,Kresse1994a,Kresse1996,Joubert1999}). Subsequently, a ``bare'' value of $\tilde{p}_{if}$ was obtained as the root-mean-square of 12 individual values. We obtained $\tilde{p}_{if}=0.013$ a.u. (atomic, or Hartree, units). Transition dipole moment is then defined as $\tilde{d}_{if}=e\tilde{p}_{if}/m\omega_{if}$, where $\omega_{if}$ corresponds to the energy difference of Kohn-Sham orbitals pertaining to $e$ states and the CBM. As presented in Figure~\ref{fig:HRfactor}, we obtain a value of $\tilde{d}_{if}=0.10$ a.u.

The ``bare'' values have to be corrected for the fact that actual calculations have been performed for the {\it negatively} charged defect, and in supercell calculations the conduction band states are repelled from the defect due to Coulomb repulsion. Once the NV center is ionized, this repulsion vanishes. Speaking strictly, we would have to evaluate matrix element between two multi-electron states, described by two separate Slater determinants: the first state corresponding to the negatively charged NV, the second state corresponding to the neutral NV plus an electron in the conduction band. The correction appears when we approximate the true matrix elements between multi-electron states with matrix elements between single-electron Kohn-Sham states.

Repulsion between the NV center and conduction band states can be taken into account via the so-called Sommerfeld factor $f$\cite{Passler1976}. $f$ quantifies the change of the density of CBM electrons near a charged defect relative to the case if the defect was neutral. We have evaluated $f$ by investigating the behavior of conduction band electrons in the presence of a negatively charged defect using a model Hamiltonian set to reproduce the behavior of the actual defect in diamond (the parameters in the model Hamiltonian are: the supercell lattice constants, charge of the defect, dielectric constant, effective mass of electrons, and the extent of the defect wave-function).  We obtain the value $f=0.4$, which yields the ``true'' values of matrix elements $p_{if}=\tilde{p}_{if}/\sqrt{f}=0.21$ a.u. and $d_{if}=\tilde{d}_{if}/\sqrt{f}=0.16$ a.u. In SI units the latter value corresponds to the transition dipole moment $d_{if}=0.085$ $e\cdot\text{\AA}$.

\section*{Section D: Electron-phonon scattering rates}
\subsection*{Surface confinement}

Consider a nanowire confining potential realized through surface confinement. The electron-phonon (e-p) scattering rate can be calculated using Fermi's golden rule for the perturbing Hamiltonian
\begin{equation}\label{HAMep2}
\hat{\mathcal{H}}_\text{ep} = \Xi_d \vec{\nabla}\cdot \vec{u}_f,
\end{equation}
where $\Xi_d$ is the deformation potential and $\vec{u}_f$ is the phonon field. Acoustic phonon modes in nanowire structures have previously been determined through elasticity theory\cite{Nishiguchi1997}. Four types of modes exist, classified as dilational, flexural, torsional and shear. However, only dilational modes possess a non-zero divergence and so contribute to a non-zero e-p scattering as per Hamiltonian~\eqref{HAMep2}. Scattering from optical modes as well as higher order processes are negligible at the liquid helium temperatures considered in this work. Following elasticity theory, the dilational modes $\vec{u}_d$ can be derived from a scalar field $\chi$ as
\begin{equation}\label{scalarPot}
\vec{u}_d=\vec{\nabla}\chi,
\end{equation}
which satisfies the wave equation
\begin{equation}\label{WE}
-\nabla^2\chi = c_l^{-2}\omega^2\chi.
\end{equation}
The precise form of $\chi$ can be determined by the normalization constraint
\begin{equation}\label{norm}
\int_V |\vec{u}_d|^2\text{d}^3 r = V,
\end{equation}
where $V$ is the volume of the wire, as well as a suitable choice of boundary conditions. For free-standing wires, we impose $\hat{n}\cdot\vec{\nabla}\chi=0$.

The solution to~\eqref{WE} can be obtained through separation of variables. For a nanowire of length $L$ and width $w$ aligned along the $\hat{z}$-axis, the resulting wavefunction is quantized by the positive integers $m=(m_x,m_y,m_z)$ and has the form
\begin{equation}\label{chi}
\chi_m = \frac{\sqrt{8}c_l}{\omega_m}\cos\left ( \frac{m_x \pi}{w}x\right )\cos\left ( \frac{m_y \pi}{w}y\right )\cos\left ( \frac{m_z \pi}{L}z\right ),
\end{equation}
with angular frequencies given by
\begin{equation}\label{AM}
\omega_m^2 = \pi^2c_l^2\left ( \frac{m_x^2+m_y^2}{w^2}+\frac{m_z^2}{L^2}\right ),
\end{equation}
where $c_l$ is the longitudinal speed of sound in diamond. The phonon field can then be written as
\begin{equation}\label{phononFieldDis}
\vec{u}_f=\sum_m \vec{u}_{d,m}\left ( \frac{\hbar}{2\rho_C V \omega_m}\right )^{1/2}\left (\hat{a}_m+\hat{a}_m^\dagger \right ),
\end{equation}
for phonon annihilation (creation) operator $\hat{a}_m$ ($\hat{a}_m^\dagger$) and $\rho_C$ the density of diamond. The scattering Hamiltonian~\eqref{HAMep2} is then calculated as
\begin{align}
\hat{\mathcal{H}}_\text{ep}=\Xi_d\vec{\nabla}\cdot \vec{u}_f&=\Xi_d\sum_m \left ( \vec{\nabla}\cdot\vec{u}_{d,m} \right )\left ( \frac{\hbar}{2\rho_C V\omega_m }\right )^{1/2}\left (\hat{a}_m+\hat{a}_m^\dagger \right )\nonumber\\
&=\Xi_d\sum_m \left ( \nabla^2\chi_m\right )\left ( \frac{\hbar}{2\rho_C V\omega_m}\right )^{1/2}\left (\hat{a}_m+\hat{a}_m^\dagger \right ) \nonumber \\
&=\frac{\Xi_d}{c_l^2}\sum_m \omega_m^2\chi_m\left ( \frac{\hbar}{2\rho_C V\omega_m}\right )^{1/2}\left (\hat{a}_m+\hat{a}_m^\dagger \right ).
\end{align}

Following Fermi's golden rule, the scattering rate can be derived as
\begin{equation}\label{G_ns}
\Gamma_\text{ep}=\frac{2\pi}{\hbar^2}\sum_{n,m}\Xi_d^2 |M_{n,m}|^2\frac{\hbar}{2\rho_C V \omega_m}n_B(\omega_m,T)\rho(\omega_n-\omega_m),
\end{equation}
where $$M_{n,m}=\int_VF_n^*\chi F_0\text{d}^3 r$$ is the overlap integral between the electronic and phonon wavefunctions and $n_B$ is the Bose-Einstein distribution. Finally, $\rho$ is the density of states which is assumed to be Lorentzian and is given by
\begin{equation}
\rho(\omega_n-\omega_m)=\frac{\gamma/\pi}{(\omega_n-\omega_m)^2+\gamma^2},
\end{equation}
where $\gamma\approx \omega_m/Q$ for $Q$ the phonon quality factor.

\subsection*{Electrostatic confinement}
Now consider a nanowire confining potential realized through electrostatic confinement with electrodes. The e-p scattering rate is therefore due to the interaction between bulk phonons and confined nanowire electronic states. As above, we proceed using Fermi's golden rule with the Hamiltonian~\eqref{HAMep2} and modify only the phonon field. Only dilational modes contribute to scattering which can be calculated using the scalar field $\chi$ and wave equation~\eqref{WE}. The solution for bulk modes is given trivially by
\begin{equation}
\chi_k=\frac{\omega}{c_l} e^{i\vec{k}\cdot\vec{r}},
\end{equation}
for a continuous wavevector $\vec{k}$ with dispersion relation $\omega=c_l |\vec{k}|$. This produces the phonon field
\begin{equation}\label{phononFieldCont}
\vec{u}_f(\vec{r})=\frac{V}{(2\pi)^3}\int \vec{u}_{d,k}\left ( \frac{\hbar}{2\rho_C V \omega_k}\right )^{1/2}\left (\hat{a}_k+\hat{a}_k^\dagger \right ) \text{d}^3k.
\end{equation}
The deformation potential is therefore
\begin{align}
\hat{\mathcal{H}}_\text{ep}=\Xi_d\vec{\nabla}\cdot \vec{u}&=\frac{\Xi_d V}{(2\pi)^3}\int \left ( \vec{\nabla}\cdot\vec{u}_{d,k} \right )\left ( \frac{\hbar}{2\rho_C V\omega_k}\right )^{1/2}\left (\hat{a}_k+\hat{a}_k^\dagger \right ) \text{d}^3 k \nonumber\\
&=\frac{\Xi_d V}{(2\pi)^3c_l^2}\int \omega_k^2 \chi_k\left ( \frac{\hbar}{2\rho_C V\omega_k}\right )^{1/2}\left (\hat{a}_k+\hat{a}_k^\dagger \right ) \text{d}^3 k.
\end{align}

Following Fermi's golden rule, the scattering rate may be derived as
\begin{align}
\Gamma_\text{ep}&=\frac{2\pi}{\hbar^2}\sum_n\frac{\Xi_d^2}{(2\pi)^3c_l^4}\int G_n(k) \omega_k^4\left ( \frac{\hbar}{2\rho_C \omega_k}\right )n_B(\omega_k,T)\delta(\omega_n-\omega_k)\text{d}^3k \nonumber\\
&=\frac{1}{2(2\pi)^2}\frac{\Xi_d^2}{\hbar\rho_C c_l^4}\sum_n \int G_n(k)\omega_k^3 n_B(\omega_k,T)\delta(\omega_n-\omega_k)\text{d}^3k,\label{GepNS}
\end{align}
where we have defined the squared magnitude of the overlap integral as
\begin{equation}\label{Gnk}
G_n(k)=\left | \int_0^w \int_0^w \int_0^L F_n^* \left (\frac{c_l}{\omega_k}e^{i\vec{k}\cdot\vec{r}} \right )F_1\text{d}x \text{d}y \text{d}z\right |^2.
\end{equation}
Expression~\eqref{GepNS} can be simplified through converting the integral over $k$-space into polar coordinates. Employing the properties of the Dirac delta distribution leads to the final scattering rate
\begin{equation}
\Gamma_\text{ep}=\frac{1}{2(2\pi)^2}\frac{\Xi_d^2}{\hbar\rho_C c_l^7}\sum_n \omega_n^5 n_B(\omega_n,T)\int_0^\pi\int_0^{2\pi}G_n\left ( \frac{\omega_n}{c_l},\theta,\phi\right )\sin(\theta)\text{d}\theta\text{d}\phi,
\end{equation}
where the angular integrals must be evaluated numerically.

\section*{Section E: Electrostatic confinement using electrodes}

\begin{figure}[h]
	\centering
	\includegraphics[width=1\textwidth]{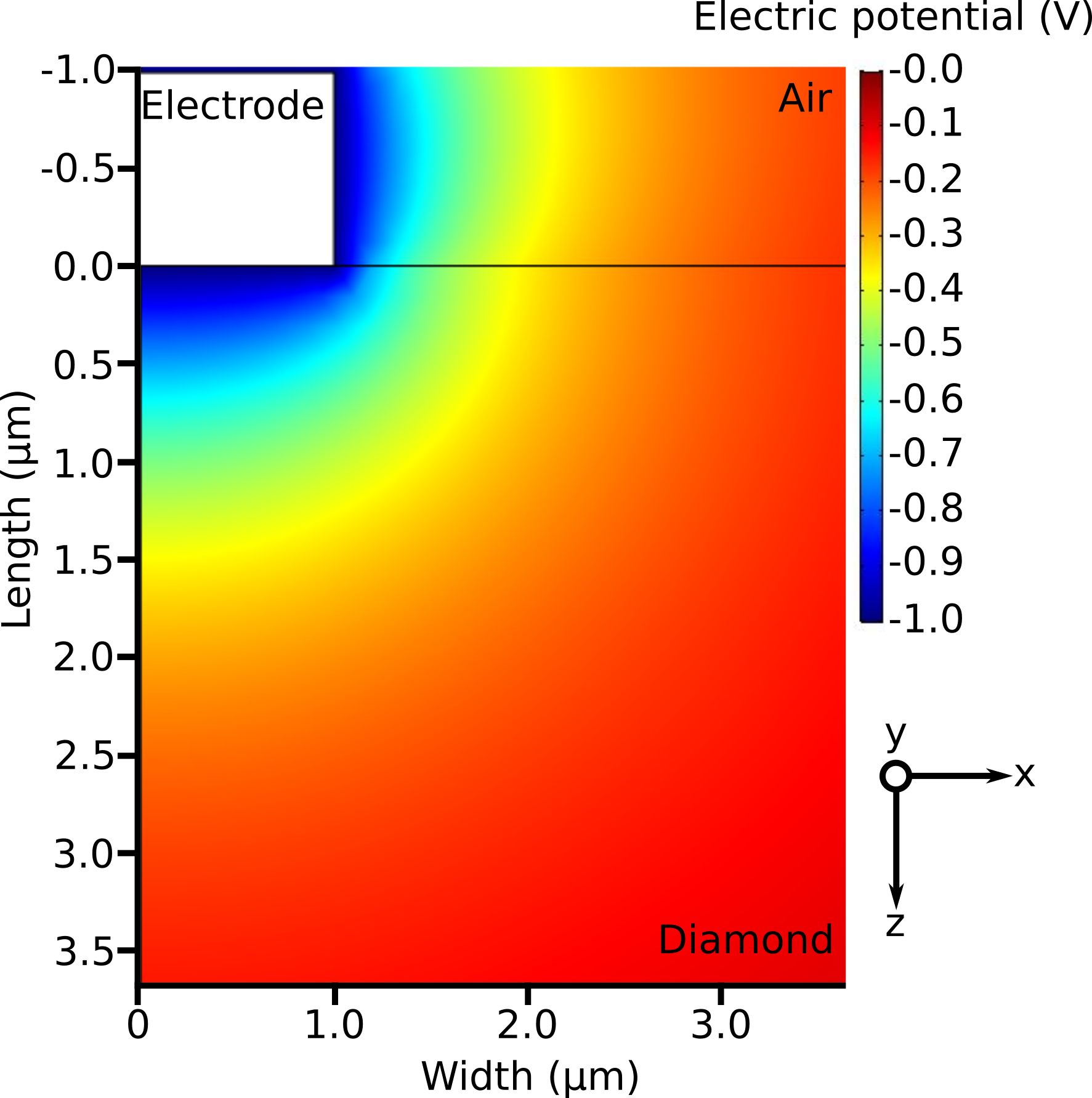}
	\caption{COMSOL Multiphysics\textregistered \ simulations of the electric potential produced by a cylindrical electrode on a diamond substrate. The substrate is 10~$\mu$m deep in the $z$ direction and extends infinitely in the $x$ and $y$ directions. Depicted here is a cross-section of the substrate in the $x$-$z$ half-plane ($y=0$) with the electrode centered at the origin (note that the simulation has rotational symmetry). A potential of $+1$ V has been applied to the electrode which has a height and radius of 1 $\mu$m. The electrostatic potential has an approximately wire-like geometry with a longitudinal axis extending into the substrate.}
	\label{fig:comsol}
\end{figure}

We propose realising a nanowire confining potential through electrostatic confinement in bulk diamond with nano-electrodes. To justify the feasibility of this design and explore its electrostatic properties, simulations of the confining potential where performed using COMSOL Multiphysics\textregistered \ version 5.3. As an example, we considered the confining potential produced by a cylindrical electrode of height and radius 1~$\mu$m positioned on top of a diamond substrate 10~$\mu$m thick. On the bottom surface of the substrate, a grounded plate electrode was simulated by enforcing that the electric potential must vanish. We find that a modest potential of $+1$~V applied to the top electrode produces a confining potential which extends micrometers deep into the substrate. This is depicted in Figure~\ref{fig:comsol}, a density plot of the electric potential within a cross-section of the diamond lattice. As can be seen, the confining potential is approximately shaped like a wire and localizes the electron density near the electrode. The effective length of this wire can be tuned by modifying the applied voltage.

\bibliographystyle{spiejour}   
\bibliography{STIRAP}